\RequirePackage{amsmath}
\documentclass[a4paper,12pt]{iopart}

\usepackage{subcaption}
\usepackage{graphicx}
\usepackage{placeins}
\usepackage[colorlinks=true, linkcolor=blue, citecolor=blue, urlcolor=blue]{hyperref}
\usepackage{algorithm2e}
\usepackage[switch]{lineno}
\usepackage{amsfonts}
\usepackage{mathtools}
\usepackage{float}

\usepackage[dcucite,abbr]{harvard}

\newcommand{\RNum}[1]{\uppercase\expandafter{\romannumeral #1\relax}}

\begin{document}

\title[]{Exploiting network topology in brain-scale simulations of spiking neural networks}

\author{Melissa Lober\,$^{1,2,*}$, Markus Diesmann\,$^{1,3,4}$, Susanne Kunkel\,$^{6}$}

\address{$^{1}$Institute for Advanced Simulation (IAS-6), Jülich Research Centre, Jülich, Germany\\
$^{2}$RWTH Aachen University, Aachen, Germany \\
$^{3}$Department of Physics, Faculty 1, RWTH Aachen University, Aachen, Germany\\
$^{4}$Department of Psychiatry, Psychotherapy and Psychosomatics, School of Medicine, RWTH Aachen University, Aachen, Germany\\
$^{5}$JARA BRAIN Institute Structure-Function Relationships (INM-10), Jülich Research Centre, Jülich, Germany\\
$^{6}$Neuromorphic Software Ecosystems (PGI-15), Jülich Research Centre, Jülich, Germany}
\ead{m.lober@fz-juelich.de}
\vspace{10pt}
\begin{indented}
\item[]June 2026
\end{indented}

\begin{abstract} 
Simulation code for conventional supercomputers serves as a reference for neuromorphic computing systems. The present bottleneck of distributed large-scale spiking neuronal network simulations is the collective communication between compute nodes. Communication speed seems limited by the interconnect between the nodes and the software library orchestrating the transfer of data packets. Profiling reveals, however, that the variability of the time required by the compute nodes between communication calls is large. The bottleneck is in fact the waiting time for the slowest node. A statistical model explains total simulation time on the basis of the collective distribution of computation times between communication calls.
A fundamental cure is to avoid communication calls because this requires fewer synchronizations and reduces the relative variability of computation times across compute nodes. The spatial organization of the mammalian brain into areas lends itself to such an optimization strategy. The connections between neurons within an area have short delays, but the delays of the long-range connections across areas are an order of magnitude longer. This suggests a structure-aware mapping of areas to compute nodes allowing for a partition into more frequent communication between nodes simulating a particular area and less frequent global communication. We demonstrate a substantial performance gain on a real-world example.
Our study proposes a local-global hybrid communication architecture for large-scale neuronal network simulations as a first step in mapping the structure of the brain to the structure of a supercomputer. We challenge the long-standing belief that the bottleneck of simulation is the synchronization inherent in the collective calls of standard communication libraries. The work provides guidelines for the energy efficient simulation of neuronal networks on conventional computing systems and raises the bar for neuromorphic systems.
\end{abstract}

\newpage
\section{Introduction}
\label{sec:Introduction}

As neuroscience uncovers the structure of the mammalian brain the field progressively demands the construction and simulation of larger neuronal network models. Already a few decades ago anatomists pointed out a dual universality. In the evolution of mammalian species the local cortical circuit exhibits a high degree of conservation, but from mice to human the volume of the brain increased by three orders of magnitude. Furthermore, in a single species, the local circuitry looks very similar independent of whether the respective part of the brain is processing visual or auditory information, or is involved in motor planning. Thus, there is an indication from neuroscience that larger brains have increased cognitive abilities and support more flexible behaviors \cite{Benson-Amram16,Herculano-Houzel17}.
In addition, theoretical work just concentrating on the dynamics of neuronal networks shows that downscaling inevitably perturbs the dynamics of neuronal networks \cite{Albada15}. While the first order statistics (e.g., spike rate) can easily be preserved, there are harsh limits already for the second order statistics (e.g., correlation of spiking activity). This is relevant because the second order statistics of neuronal activity directly interacts with mechanisms of plasticity (e.g., STDP; \citename{Bi98}, \citeyear*{Bi98}) and learning. However, no theoretical framework is yet available explaining the expansion of capabilities of biological brains with size. Due to this situation, until a few years ago researchers working either experimentally or theoretically on the mammalian brain were regularly asked in scientific reviews why in the face of the technical difficulties, they are not first studying the brains of arthropods (insects) or nematodes (worms). There are good reasons to study small brains for conceptual and applied reasons as no mammal can fly like a fly. But since the rise of large-scale artificial neural networks, the criticism on the study of large brains has faded. From this field comes the evidence that large-scale networks have qualitatively different capabilities \cite{Berti25}.

Next to the functional considerations, also brain structure can only faithfully be represented in large models. The mammalian cortex unfolds to an, in some approximation, isotropic sheet, in the human covering a quarter of a square meter with a thickness only in the millimeter range \cite{toro08_2352,Fischl00_11050}. The vertical dimension exhibits a layered structure, constituting the cortical microcircuit mentioned above. The horizontal dimensions are dominated by a species-dependent space constant of a few hundred micrometers for the connection probability between pairs of neurons \cite{Stepanyants2008_13,Boucsein11_1,Kurth25}. The probability that two randomly selected neurons are connected is low, reflecting pronounced sparsity in space. Certain species like carnivore animals or primates additionally exhibit a patchy super-structure establishing connections between farther apart neurons \cite{Voges10_137}. This network architecture suggests a scale invariant network activity once model size sufficiently exceeds the space constant. However, the mammalian cortex decomposes into around a hundred areas (again, specifics are species dependent), and below the cortical sheet run long-range shortcuts directly connecting them. These long-range connections organize the functional specificity. Thus, even when we leave out the functionally important loops of cortex with subcortical components like thalamus, basal ganglia, and hippocampus, cortex itself represents a hierarchical recurrent network with at least three levels of organization. This imposes further constraints on a meaningful downscaling. The sparsity of the mammalian brain is spatio-temporal; if there is a connection between two neurons, it mediates only a few spikes per second. And there is evidence that even this number is overestimated as experimental procedures have a bias towards more active neurons \cite{Barth12}.

Long-range cortico-cortical connections span substantially larger anatomical distances than local intra-area connections and therefore exhibit larger conduction delays. Experimental measurements of response latencies to electrical stimulation report a mean axonal conduction velocity of approximately $3.5\,\text{m}/\text{s}$ for projections between visual areas in macaque monkey \cite{Girard01}. Given that inter-area distances can reach several tens of millimeters \cite{Schmidt18_1409}, the resulting spike transmission delays are on the order of several milliseconds. While local connections may exhibit a broad distribution of delays that can extend into the millisecond range, their shortest delays typically remain well below those of long-range projections. This separation of temporal scales between local and long-range interactions provides the motivation for the partitioning and communication scheme introduced in this work.

The idea of mapping the structural organization of neural systems onto computing architectures has a long history in computational neuroscience and high-performance simulation. Early work on parallel discrete-event simulation established graph partitioning as a central tool for distributing interacting entities across processors while minimizing communication overhead \cite{Kernighan1970,Fujimoto90}. Such approaches were adopted in neural simulations primarily when connectivity was irregular or network structure had to be inferred from an instantiated graph. More recently, these concepts have also been applied to spiking neural networks, for example through hypergraph-partitioning–based neuron allocation strategies \cite{Fernandez-Musoles19}. More recent simulation frameworks extend this approach with CORTEX employing indegree sub-graph decomposition of instantiated network graphs to enable highly parallel execution \cite{Lyu24}, and STACS using graph-based partitioning to optimize spike communication for spatially structured networks \cite{Wang_2024}.
In parallel, several structure-aware approaches exploit explicit properties of cortical organization. Tiling-based methods decompose neural tissue into spatial blocks, motivated by the locality of cortical connectivity, thereby reducing communication to neighboring tiles \cite{Kozloski11_5_15,yamaura20_16}. Earlier still, the SPLIT simulator demonstrated that exploiting explicit population structure and synaptic delays yields efficient mappings of neural components to distributed hardware \cite{Hammarlund98_443}. Another early example in this direction is the modular, event-driven framework NEOSIM \cite{Goddard01}, which points out the role of axonal delays in organizing parallel execution. While these approaches are often tailored to specific models or architectures, they highlight that many large-scale brain models define modular organization and connectivity rules explicitly. Unlike graph partitioning approaches, which aim to uncover the hidden structure of an instantiated graph, these works exploit the fact that the modular organization of brain models is known a priori from the underlying biology and is directly reflected in the instantiation rules used during network construction. The present work adopts the same perspective, and extends it by exploiting not only the modular organization of the network but also the temporal structure imposed by synaptic transmission delays to reduce the frequency of communication.
Many works in the field of neuromorphic computing pursue these principles found a quarter of a century ago, for example through specialized hardware architectures. For neuromorphic hardware systems such as BrainScaleS \cite{Pehle22} and SpiNNaker \cite{Rhodes19_20190160}, where resource constraints inherently limit the size and type of networks that can be simulated, efficiently mapping network structure onto hardware resources is crucial for achieving good performance. Extending this perspective to conventional high performance computing (HPC) systems is the starting point of the present work.

With the field of dedicated hardware for neuronal simulations growing rapidly over the past years, conventional HPC platforms continue to serve as an essential reference benchmark, offering greater flexibility at potentially lower cost. Work by \citeasnoun{Kurth22} demonstrates that, for a biologically realistic cortical microcircuit model, conventional HPC-based neuronal simulations keep pace with specialized neuromorphic systems such as SpiNNaker \cite{Rhodes19_20190160} in terms of both real-time performance and energy efficiency (see \citename{Senk26} \citeyear*{Senk26} for a review). These results highlight the importance of further developing methods to exploit the full potential of conventional hardware in large-scale neuronal network simulations.

This history of three decades of investigation on the efficient partitioning of the graph representing a neuronal network begs the question why none of the results is realized in production simulation code (set aside the general difficulty to operate scientific software as infrastructure;
\citename{Gewaltig14} \citeyear*{Gewaltig14},
\citename{Hocquet24_1} \citeyear*{Hocquet24_1},
\citename{Plesser25_295} \citeyear*{Plesser25_295}
). Initially the computational load of propagating the dynamical state of the neurons and synapses and the effort in routing the spikes on a compute node was just so high that communication was not the bottleneck. Furthermore, the past two decades have been dominated by the study of essentially random local networks where each neuron has some probability to establish a synapse with any other neuron. It is only now that network models have become large enough to expose the non-local hierarchical architecture of the brain. The combination of these two factors, powerful compute nodes and large structured networks, entices us to revisit the idea of effectively mapping the structure of the neuronal network onto the structure of the computing system. The idea is not to derive an algorithm for the most general connectivity graph, but a scheme that provides advantage for the brain models presently under study.

Distributed large-scale spiking network simulations on HPC systems use the MPI standard \cite{mpi40}, which enables multiple independent processes to communicate across compute nodes in a distributed-memory environment. Each MPI process (independent parallel worker with separate memory space) is responsible for a subset of neurons, allowing the simulation to scale beyond the resources of a single compute node. Each MPI process employs multi-threading to harness all cores of a compute node efficiently. Such simulations typically advance in discrete time steps and require periodic global exchange of spikes to preserve causality. The global minimum synaptic transmission delay $d_\text{min}$ determines how far the simulation can progress independently on each MPI process before communication is required \cite{Morrison05a}. We refer to this interval with fixed length $d_\text{min}$ as a \emph{simulation cycle} (\fref{fig:algorithm}). Each cycle consists of three phases: \emph{deliver}, which applies incoming spikes to local neurons; \emph{update}, which advances neuronal states and generates spikes; and \emph{collocate}, which collects emitted spikes for communication. At the end of each cycle, all processes exchange spikes via a collective MPI operation in the \emph{communicate} phase.

Over the past decade, the dominant performance bottleneck of the simulation cycle has shifted repeatedly as simulation technology evolved. Early versions of code for the distributed simulation of spiking neuronal networks employed \texttt{MPI\_Allgather()}, causing spike communication to dominate runtime \cite{Lansner12_283}. The introduction of directed spike communication based on \texttt{MPI\_Alltoall()} --- a collective MPI operation in which each process sends a distinct message to every other process --- substantially reduced communication costs, and spike delivery emerged as the dominant bottleneck across a wide range of network sizes \cite{Jordan18_2}. Subsequent optimizations targeting spike delivery \cite{Pronold22_102952,pronold2022} again shifted the balance, rendering communication a limiting factor.

\begin{figure}
    \includegraphics[scale=1]{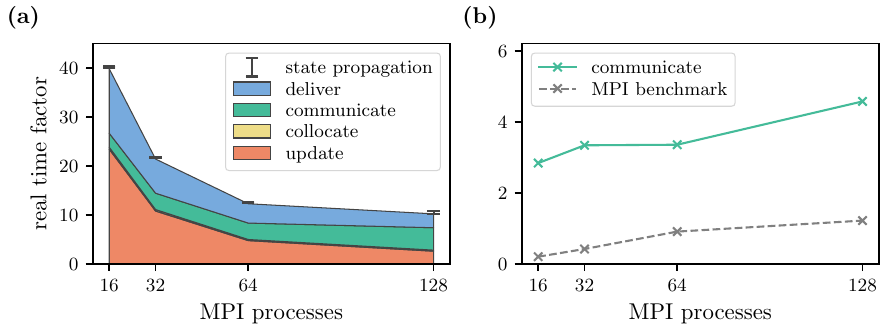}
    \caption{\textbf{Strong-scaling of the multi-area model of macaque visual cortex (MAM) in the ground state}. (a) Real time factors, defined as wall-clock time normalized by simulated model time, stacked for each phase of the simulation (legend). All simulations cover $T_\mathrm{model}=10\,\mathrm{s}$ of biological time. Error bars (at line resolution) indicate variability across three simulations using different random seeds. (b) Real time factor of communication phase extracted from panel (a) including synchronization. Dashed curve marks time attributed to pure MPI communication, as estimated from MPI benchmarks (see \fref{fig:mpi_benchmark}) for a $T_\mathrm{model}=10\,\mathrm{s}$ simulation and average buffer sizes per target rank of $1408$, $837$, $514$, $317$ bytes reported by the simulations in (a) using $16, 32, 64, 128$ MPI processes, respectively. All data obtained on SuperMUC-NG using NEST version $3.6$.}
    \label{fig:MAM_and_mpi_benchmark}
\end{figure}

\Fref{fig:MAM_and_mpi_benchmark}a reports the strong-scaling performance of all mentioned simulation phases on the example of a multi-area model of macaque visual cortex (MAM; \citename{Schmidt18_e1006359} \citeyear*{Schmidt18_e1006359},
\citename{Schmidt18_1409} \citeyear*{Schmidt18_1409}%
). While simulation phases such as neuronal update and spike delivery scale reasonably well with the number of compute nodes, as expected, the graph reveals spike communication as the primary bottleneck. A common assumption is that the communication cost arises mainly from the volume of data exchanged between nodes or by network latency, i.e.\ the time until the first byte of a message arrives. However, standard MPI benchmarks (\fref{fig:MAM_and_mpi_benchmark}b) show that both, the pure data transfer time and the intrinsic latency of the collective communication, are substantially lower than the communication time observed in the actual large-scale network simulations. This discrepancy reveals that neither bandwidth limitations nor MPI latency are the primary bottleneck in our case. Instead, we find that the dominant contribution to spike-communication time is synchronization overhead. This originates from the variability across compute nodes in completing the simulation cycle. Already \citeasnoun{Fernandez-Musoles19} point out synchronization as one of the key performance bottlenecks of distributed large-scale spiking neuronal networks. These findings suggest that considerable performance headroom remains untapped, and that optimizing communication schemes in distributed neuronal simulations is both necessary and feasible. 

In this work, we present a novel approach that rethinks neuron distribution and communication between compute nodes. By reducing synchronization overhead, we achieve a substantial reduction in communication time. Furthermore, we show that these improvements extend beyond communication itself, also accelerating subsequent phases such as spike delivery.

%
The presented conceptual and algorithmic work is part of our long-term collaborative project to provide the technology for neural systems simulations \cite{Gewaltig_07_11204}.
Preliminary results have been presented in abstract form \cite{Lober24-FENS,Lober24-ICNCE}.

\FloatBarrier
\section{Results}
\label{sec:Results}
On conventional hardware, simulations of large-scale recurrent spiking networks with bio-plausible synaptic densities and transmission delays are bottlenecked by spike routing (\fref{fig:MAM_and_mpi_benchmark}). The routing in such distributed simulations requires, first, the exchange of spikes among all compute nodes and, second, the local delivery of spikes to their target synapses and neurons. In the case that spikes are exchanged using blocking collective MPI communication, each MPI call introduces a synchronization barrier. This entails waiting times due to variations in processing times between MPI processes. To address this bottleneck, we introduce a structure-aware communication scheme that exploits the separation of delay scales between intra- and inter-area connections. By decoupling fast local interactions from slower long-range communication, the proposed approach reduces synchronization frequency without affecting the temporal resolution of spikes.
Spike delivery is inherently irregular: it is unpredictable which neurons emit spikes in any given simulation cycle, and the local target synapses of any two neurons typically reside in different memory locations. The local target synapses of a single neuron, however, are laid out consecutively in memory, allowing a spike to reach several targets in a row. Previous work established that cache efficiency in the spike-delivery phase depends critically on the number of target synapses each neuron has on the local process and thread: the more local targets, the better the cache reuse \cite{Pronold22_102952,pronold2022}. When scaling up simulations, targets are increasingly spread out, and each spike finds ever fewer local targets per process and thread. As a consequence target synapses are increasingly likely to lie in uncached memory, resulting in irregular memory access. To recover cache-efficiency, the new structure-aware scheme aggregates intra-area synapses on a single compute node, thereby increasing target locality.

This study investigates to what extent a structure-aware approach alleviates these bottlenecks in simulations of multi-area models. \Sref{sec:structure-aware-simulation-strategy} introduces the proposed simulation strategy. \Sref{sec:synch-time-model} and \sref{sec:memory-access-model} develop theoretical models of why the structure-aware approach reduces synchronization times and spike-delivery times, respectively. 
These models rest on simplifying assumptions that are not fully satisfied in practice and should therefore be understood as providing qualitative mechanistic insight into the expected performance gains rather than exact quantitative predictions.
Finally, \sref{sec:Benchmark-results} presents benchmarking results obtained with our reference simulation code NEST comparing the conventional and the structure-aware simulation strategy.

\subsection{Structure-aware simulation strategy}
\label{sec:structure-aware-simulation-strategy}
\begin{figure}
  \centering
  \includegraphics[width=0.9\linewidth]{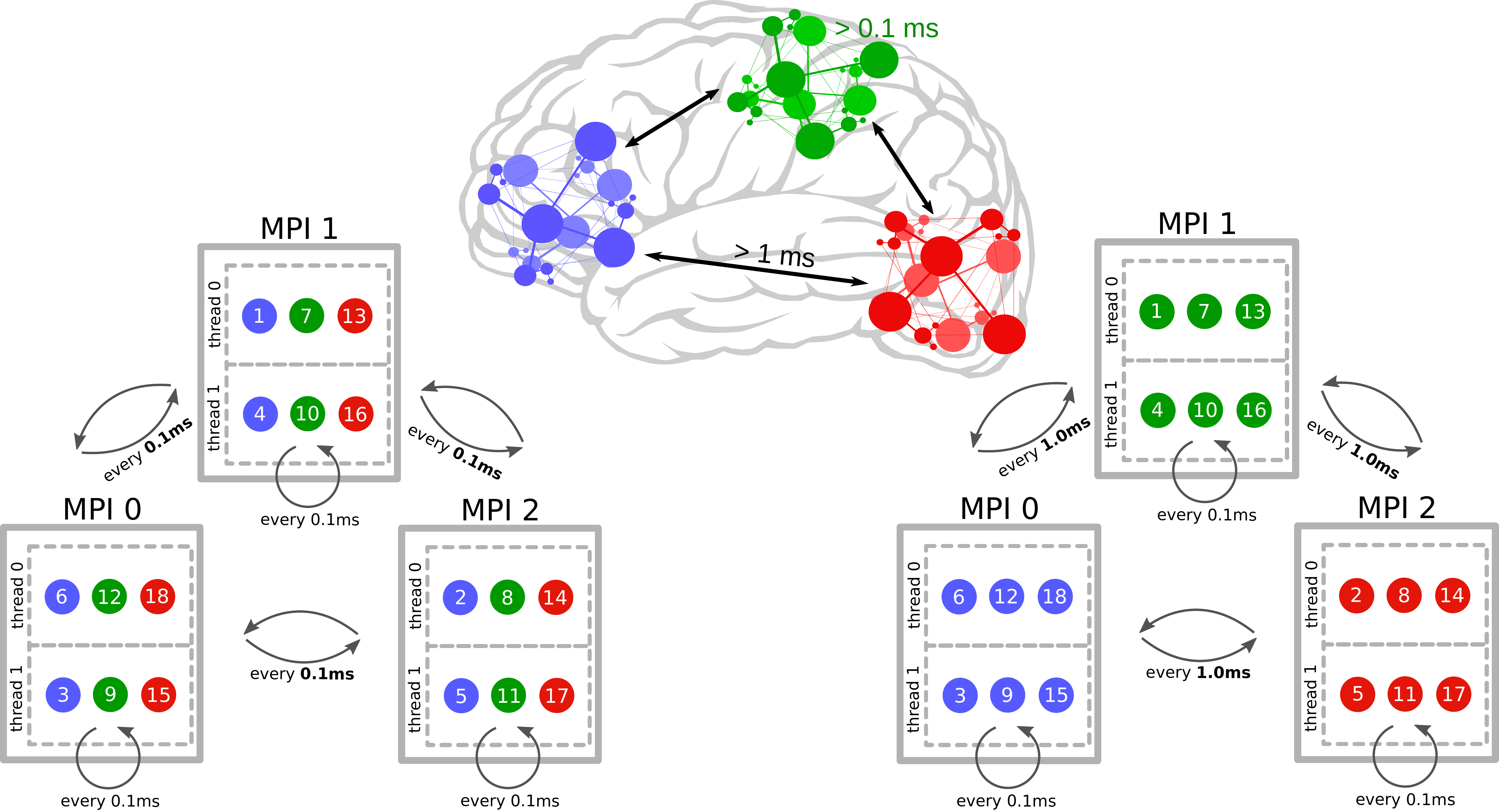}
  \caption{\textbf{Conventional and structure-aware simulation scheme illustrated for an example multi-area model.} Neurons of three areas (middle) are color coded in blue, red, and green. Synaptic transmission delays are significantly shorter within areas than between areas; here, the minimum delays are $0.1\,\text{ms}$ and $1.0\,\text{ms}$, respectively. The minimum delay between any pair of neurons represented on two different MPI processes dictates the communication interval for the exchange of spikes between the processes. In conventional simulation technology, neurons are distributed across MPI processes and threads according to a round-robin scheme (left) to balance workload. Therefore, network structure can not be exploited, and global MPI communication is required every $0.1\,\text{ms}$. The structure-aware distribution scheme (right) maps areas to MPI processes, increasing the required interval for global communication to $1.0\,\text{ms}$. Brain outline in background by Gordon Johnson via Pixabay (\url{https://pixabay.com/}).}
  \label{fig:morph}
\end{figure}
Conventionally, simulation codes balance computational workload by distributing neurons and their incoming synapses across MPI processes and threads in a round-robin fashion disregarding neuronal populations \cite{Morrison05a,Migliore06_119}. For the specific case of a multi-area simulation, this structure-agnostic approach entails that the short synaptic transmission delays within areas enforce frequent global MPI communication in order to maintain causality (\fref{fig:morph}).
The structure-aware distribution strategy, in contrast, maps areas to MPI processes. This allows for a reduction in global MPI communication taking advantage of the minimum inter-area synaptic delay $d^\mathrm{inter}_\mathrm{min}$ being significantly longer than the overall minimum delay $d_\mathrm{min}$. Henceforth, we constrain inter-area delays in such a way that $d^\mathrm{inter}_\mathrm{min}$ is a multiple of $d_\mathrm{min}$ and let
\begin{equation}
    D\,=\,\frac{d^\mathrm{inter}_\mathrm{min}}{d_\mathrm{min}}
    \label{eq:d-min-ratio}
\end{equation}
define the integer ratio between these two minimum delays.

\subsubsection{Simulation flow}
\begin{figure}
    \centering
    \includegraphics[scale=1]{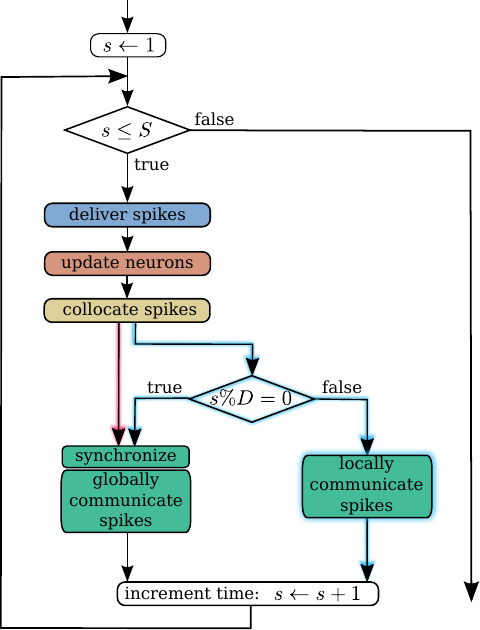}
    \caption{\textbf{Conventional and structure-aware simulation flow.} Flow chart of the iteration over $S$ simulation cycles highlighting differences between the conventional simulation strategy (pink shaded arrow) and the structure-aware strategy (cyan shaded arrows). For every simulation cycle, each MPI process first delivers incoming spikes from the MPI receive buffer to their local target neurons (blue box). Second, each process updates all process-local neurons (red box). And third, it collocates new spikes in the MPI send buffer (yellow box). In the conventional simulation scheme, every simulation cycle terminates with a global MPI communication of spikes (left green box). In the structure-aware scheme, most cycles terminate with a process-local exchange of spikes (right green box), and only every $D$-th cycle terminates with a global MPI communication.}
    \label{fig:algorithm}
\end{figure}

To evolve a model network for biological time $T_\mathrm{model}$, the simulation needs to go through $S\,=\,T_\mathrm{model}/d_\mathrm{min}$ simulation cycles (\fref{fig:algorithm}), where we assume $T_\mathrm{model}$ to be a multiple of $d_\mathrm{min}$. 
Each simulation cycle consists of three consecutive phases:
\begin{description}
    \item[Deliver] Each MPI process delivers spikes from its receive buffer to the corresponding local target neurons.
    \item[Update] Each MPI process propagates the state of its local neurons, which includes the detection of threshold crossings and where applicable, the buffering of emitted spikes in the process-local spike register \cite{Plesser07_672}. The temporary buffering enables the operation of threads free of barriers (see figure S\RNum{1}1 in \citename{Vogelsang26_arxiv}, \citeyear*{Vogelsang26_arxiv}). 
    \item[Collocate] Each MPI process transfers newly emitted spikes from the local spike register to the MPI send buffer and prepares them for communication with other processes.
\end{description}
In conventional simulation technology, all MPI processes globally communicate spikes after every simulation cycle $s$. In case of a blocking collective MPI communication, this requires all MPI processes to synchronize. The structure-aware simulation strategy, however, only requires global communication after every $D$-th cycle, and a local exchange of spikes suffices for all other cycles. Batching long-range spikes over $D$ cycles does not affect the temporal resolution of spike transmission or the resulting network dynamics. Each spike is placed into the neuron-local ring buffer (\sref{sec:Reference-implementation}) of its target neuron with the correct remaining transmission delay upon arrival, ensuring that it contributes to state propagation at precisely the correct point in simulated time regardless of when it was communicated between processes.
The local communication is a technicality left to the MPI library; it needs to efficiently swap local send and receive buffer. In the following, communication comprises both data exchange and synchronization overhead unless stated otherwise. For both simulation strategies, after every spike exchange, a new simulation cycle starts, and each process delivers incoming spikes from the MPI receive buffer to their local target neurons.

\subsubsection{Communication cost and scaling}
\begin{figure}
    \centering
    \includegraphics[scale=1]{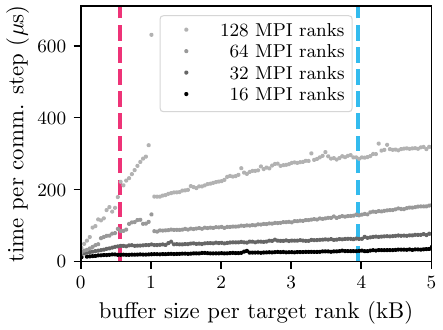}
    \caption{\textbf{MPI collective performance for increasing message sizes.} Time required for a single \texttt{MPI\_Alltoall()} call (average over $1000$ calls) as a function of buffer size when using the OpenMPI library on SuperMUC-NG, shown for increasing numbers of MPI processes as indicated in legend. Dashed vertical lines indicate the typical buffer sizes per target rank for the MAM-benchmark (see \sref{sec:benchmark-network-model}), with $\sim{}130,000$ neurons per MPI rank, simulated with the conventional (pink) or structure-aware (cyan) strategy.}
    \label{fig:mpi_benchmark}
\end{figure}
Communicating spikes only every $D$-th cycle in the structure-aware strategy implies communicating $D$-times more data at once. However, collective MPI communication benchmarks on SuperMUC-NG (\fref{fig:mpi_benchmark}) demonstrate that the time required for a single \texttt{MPI\_Alltoall()} operation scales sublinearly with the message size in the relevant range. As a result, aggregating spikes over multiple simulation cycles and sending them in a single communication step is expected to lead to a net reduction in communication time despite the increased data volume per call. For example, for $128$ MPI processes and assuming a delay ratio of $D=10$, the benchmarks predict a reduction in data-exchange time by $86\,\%$. The distinct jumps occurring for $64$ and $128$ MPI processes likely reflect switches within the OpenMPI library \cite{Gabriel04_OpenMPI} between different collective communication algorithms. 

Besides reducing global MPI communication, the structure-aware distribution scheme aggregates all synapses between neurons of the same area on a single MPI process and thereby effectively increases the average number of local target synapses per neuron, which enables a more cache-efficient delivery of spikes.

\Sref{sec:Reference-implementation} describes implementation details of our reference code NEST and the necessary changes to the code enabling the structure-aware simulation strategy. \Sref{sec:Benchmark-results} provides benchmarking results of multi-area model simulations with NEST for both the conventional and the structure-aware strategy, in particular also the contributions of deliver, update, collocate, communicate, and synchronize to the overall runtime. Let us now explore the stochastic nature of the synchronization problem.

\subsection{Theoretical analysis of synchronization}
\label{sec:synch-time-model}
\begin{figure}
    \centering
    \includegraphics[width=0.9\linewidth]{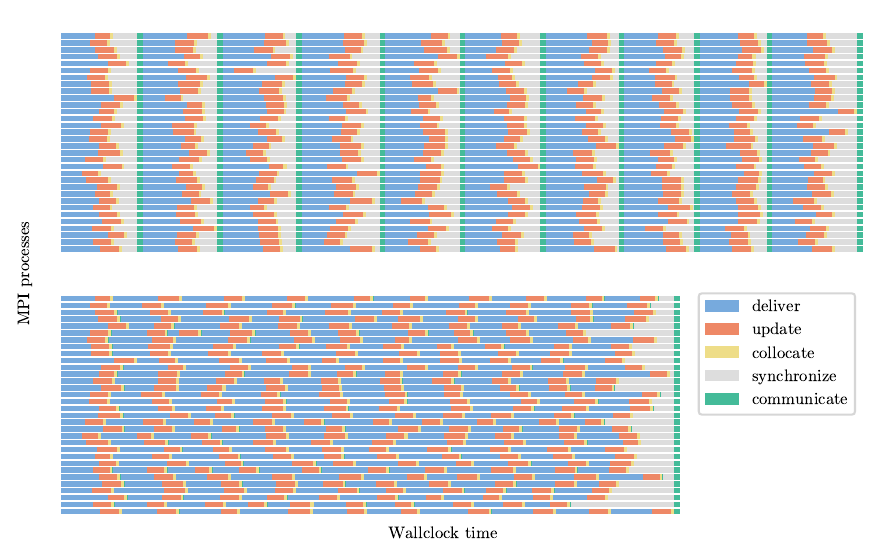}
    \caption{\textbf{Graphical intuition of the predicted reduction in overall synchronization time and thus overall runtime in a multi-area model simulation.} Illustration of $S\,=\,10$ simulation cycles on $M\,=\,32$ MPI processes when using the structure-aware simulation strategy (bottom) instead of the conventional strategy (top). All timing data is artificial and generated for illustration purposes. Wallclock time spent per simulation cycle is color coded: deliver (blue bars), update (red bars), and collocate (yellow bars). The exchange of spike data between MPI processes (communicate) is assumed to be taking up minimal time (green bars). The conventional strategy requires global communication after every simulation cycle. In case of collective blocking MPI communication, this entails synchronization (gray bars), which means that the process taking longest for the cycle requires all other processes to wait. The structure-aware strategy requires the same wallclock times per simulation cycle but reduces the number of global communications by a factor of $D$ (\eref{eq:d-min-ratio}); here we assume $D\,=\,10$. This allows for the $10$ cycles to be simulated without intermittent synchronization and thus levels out variations such that the overall synchronization times are lower than the sums of the corresponding per-cycle synchronization times in the conventional case.}
    \label{fig:sim-cycle_times_intuition}
\end{figure}

The structure-aware simulation strategy reduces the number of required global MPI communications in distributed simulations of multi-area models by a factor of $D$ (see \eref{eq:d-min-ratio} for definition). Correspondingly, in case the collective communication is blocking, MPI processes also synchronize $D$ times less often. This reduction in the number of required synchronizations potentially decreases overall synchronization time and thus runtime (\fref{fig:sim-cycle_times_intuition}), which in the following we explain with the help of a theoretical model.
\begin{figure}
    \centering
    \includegraphics[width=0.9\linewidth]{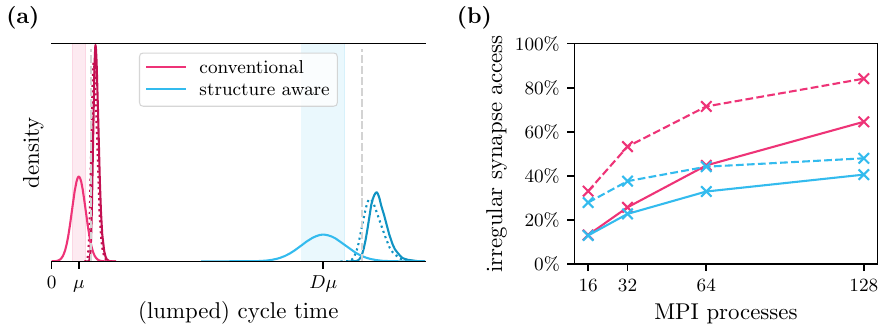}
    \caption{\textbf{Theoretical analysis explaining the advantage of the structure-aware strategy over the conventional strategy in simulations of multi-area models.} (a) Probability distributions underlying the theoretical model of synchronization times. Distributions of cycle times for the conventional strategy (pink curve) and lumped cycle times for the structure-aware strategy (cyan curve) with standard deviations indicated by light pink/cyan bands, as well as respective distributions of maxima per (lumped) cycle assuming $M\,=\,128$ (solid dark pink/cyan) and $M\,=\,64$ (dotted) MPI processes. Dashed vertical lines indicate the upper $3.5\,\%$ quantile of the cycle time distributions for $M\,=\,128$ processes, respectively. (b) Predicted fraction of irregular memory access to synapses in the spike-delivery phase in simulations of multi-area models as a function of number of MPI processes for the conventional (pink) and the structure-aware strategy (cyan) and for $T_\mathrm{M}=48$ (solid lines) and $T_\mathrm{M}=128$ (dashed lines) threads per MPI process. Weak-scaling scenario assuming $N_M\approx130,000$ neurons per process and $K_N\approx6000$ synapses per neuron. Equally sized areas of $N_M$ neurons and equal amounts of $K_N^\mathrm{intra}=K_N^\mathrm{inter}\approx3000$ intra-area and inter-area synapses are assumed for the structure-aware case.}
    \label{fig:sim-cycle_times_theory_andn_mem_access}
\end{figure}
For all $m \in M$ MPI processes and $s \in S$ simulation cycles, we assume that cycle times are governed by a normal distribution (\fref{fig:sim-cycle_times_theory_andn_mem_access}a) with mean $\mu$ and standard deviation $\sigma$:
\begin{equation}
    t_{m,s} \sim \mathcal{N} \left(\mu ,\sigma^2 \right).
    \label{eq:normal-dist-cycle-times}
\end{equation}
The conventional simulation strategy requires synchronization after every cycle. Disregarding data exchange, the total runtime therefore amounts to the sum of maximum cycle times over all $S$ cycles:
\begin{equation}
    T^\mathrm{conv}_\mathrm{wall} = \sum_{s=1}^{S} \text{max} \left\{ t_{m,s} \right\}_{m=1}^M.
    \label{eq:T^conv_wall}
\end{equation}
Approximations of order statistics and more specifically expected maxima for different sample sizes are available in the literature, for example, \citeasnoun{Blom1958_thesis} introduces an approximation for small sample sizes. With increasing sample size drawing values from the tails of the normal distribution becomes more likely, and hence the expected maximum increases. Applied to the runtime estimate of \eref{eq:T^conv_wall}, this means that with increasing number of MPI processes $M$, we expect larger maximum cycle times (\fref{fig:sim-cycle_times_theory_andn_mem_access}a) and thus an increase in overall runtime.

When using a structure-aware simulation strategy, each MPI process goes through $D$ cycles and only after every $D$-th cycle all processes synchronize. The overall runtime amounts to
\begin{equation}
    T^\mathrm{struc}_\mathrm{wall} = \sum_{l=1}^{S/D} \text{max} \left\{ t_{m,l} \right\}_{m=1}^M,
\end{equation}
where
\begin{equation}
    t_{m,l} = \sum_{s=(l-1)D+1}^{lD} t_{m,s}.
\end{equation}
denotes the lumped cycle times. Throughout this study, we use the terms \emph{cycle times} and \emph{lumped cycle times} interchangeably in the context of the structure-aware strategy to refer to $t_{m,l}$.
According to the central limit theorem (CLT), mean and variance of the initial normal distribution (\eref{eq:normal-dist-cycle-times}) are scaled by a factor of $D$ such that
\begin{equation}
    t_{m,l} \sim \mathcal{N} \left(D\mu ,D\sigma^2 \right).
    \label{eq:normal-dist-lumped-cycle-times}
\end{equation}
The CLT argument applies only under the independence assumption which \sref{sec:weak-scaling-results} shows to be violated due to persistent serial correlations in per-process cycle times.
Under this independence assumption, the lumped cycle times are expected to be $D$ times larger than the original cycle times but only with a $\sqrt{D}$ times larger standard deviation (\fref{fig:sim-cycle_times_theory_andn_mem_access}a). The ratio of the coefficients of variation of the two distributions
\begin{equation}
    \frac{CV^\mathrm{struc}}{CV^\mathrm{conv}} = \frac{1}{\sqrt{D}}
    \label{eq:rel_dispersion}
\end{equation}
quantifies that the structure-aware strategy reduces relative dispersion from the mean by $1/\sqrt{D}$. This directly translates into an expected decrease in overall synchronization time by this factor in the following way. The expected overall runtimes can be expressed as
\begin{align}
    \mathbb{E}[T^\mathrm{conv}_\mathrm{wall}]   &= \sum_{s=1}^{S} \mathbb{E}[\text{max} \left\{ t_{m,s} \right\}_{m=1}^M] \notag \\
                                                &= \sum_{s=1}^{S} \mu + \xi_M \sigma \notag \\
                                                &= S \mu + S \xi_M \sigma, \label{eq:E_of_T^conv_wall}
\end{align}
in the conventional case and
\begin{align}
    \mathbb{E}[T^\mathrm{struc}_\mathrm{wall}]  &= \sum_{l=1}^{S/D} \mathbb{E}[\text{max} \left\{ t_{m,l} \right\}_{m=1}^M] \notag \\
                                                &= \sum_{l=1}^{S/D} D \mu + \xi_M \sqrt{D} \sigma \notag \\
                                                &= S \mu + S \xi_M \frac{1}{\sqrt{D}} \sigma \label{eq:E_of_T^struc_wall}
\end{align}
in the structure-aware case. In both cases, the expected maximum is defined by the factor $\xi_M$, which indicates how many standard deviations away from the mean it is located. The factor $\xi_M$ depends on the number of MPI processes $M$ and can be approximated (e.g., \citename{Blom1958_thesis}, \citeyear*{Blom1958_thesis}). The left summand $S\mu$ in \eref{eq:E_of_T^conv_wall} and \eref{eq:E_of_T^struc_wall} corresponds to the expected contributions of deliver, update, and collocate to the overall runtime. These contributions cancel out in the difference of the expected runtimes leaving the right summands: 
\begin{equation}
    \mathbb{E}[T^\mathrm{conv}_\mathrm{wall}]-\mathbb{E}[T^\mathrm{struc}_\mathrm{wall}] = S \xi_M \sigma - S \xi_M \frac{1}{\sqrt{D}} \sigma,
\end{equation}
which correspond to the contribution of synchronize to the overall runtimes. The ratio of the expected overall synchronization times is just
\begin{align}
    \frac{\mathbb{E}[T^\mathrm{struc}_\mathrm{synch}]}{\mathbb{E}[T^\mathrm{conv}_\mathrm{synch}]} &= \frac{S \xi_M \frac{1}{\sqrt{D}} \sigma}{S \xi_M \sigma} \notag \\
    &= \frac{1}{\sqrt{D}}.\label{eq:theoretical_ratio_synch_times}
\end{align}
This theoretical prediction provides a first idea of the savings in synchronization time achievable with a structure-aware simulation strategy. In particular, since $\frac{1}{\sqrt{D}}$ declines rapidly, we conclude that the structure-aware approach is already effective for small ratios $D$ of minimum inter-area synaptic delay to overall minimum delay (\eref{eq:d-min-ratio}), while little more can be gained by further increasing $D$.
This derivation assumes fully independent consecutive cycle times. In the opposite limiting case of full correlation between all $D$ consecutive cycle times of a given MPI process, the lumped cycle time $Dt$ has standard deviation $D\sigma$ (law of functions of a random variable, \cite{PapoulisProb4th}) and thus the same coefficient of variation as the original distribution, yielding a ratio of $\mathbb{E}[T^\mathrm{struc}_\mathrm{synch}]/\mathbb{E}[T^\mathrm{conv}_\mathrm{synch}] = 1$.

For concrete cycle-time measurements obtained in benchmarking simulations (\sref{sec:Benchmark-results}), it is possible to estimate the expected interval of the per-cycle maxima. To this end, we let $p_{[q,+\infty)}$ define the probability for a cycle time to fall within the interval $[q,+\infty)$ of the cycle-time distribution. For each simulation cycle, the probability of drawing the maximum cycle time from this interval is
\begin{equation}
    p_{[q,+\infty)}^\mathrm{max} = 1 - \left( 1-p_{[q,+\infty)} \right)^M
    \label{eq:prediction-per-cycle-maximum-interval}
\end{equation}
given a specific number of $M$ MPI processes. For a simulation with $M=128$ MPI processes, this means, for example, that the upper $3.5\,\%$ of the cycle-time distribution contribute approximately to the upper $99\,\%$ of the resulting distribution of per-cycle maxima. Or vice versa, the upper $99\,\%$ of the per-cycle maxima are expected to fall within the interval containing the upper $3.5\,\%$ of measured cycle times (\fref{fig:sim-cycle_times_theory_andn_mem_access}a).

\subsection{Theoretical analysis of spike delivery}
\label{sec:memory-access-model}
Delivering a spike to its first target synapse on a specific MPI process and thread typically entails access to a memory location that is not yet cached, whereas subsequent target synapses are then sequentially accessed. The following theoretical analysis allows us to predict the fraction of first-synapse accesses in multi-area model simulations for the conventional and the structure-aware strategy.

We consider multi-area networks of $N$ neurons, each neuron with an average number of $K_N$ incoming synapses. We assume that the simulation is carried out using $M$ MPI processes, each running $T_M$ threads, resulting in $T=M\,T_M$ threads in total. $N_M=N/M$ and $N_T=N/T$ denote the process-local and thread-local number of neurons, respectively.

Conventionally, neurons are distributed across processes and threads in a round-robin fashion. The probability that a neuron has at least one target on a specific thread is given by
\begin{equation}
    p_\mathrm{target} = 1 - (1 - 1 / N)^{N_TK_N}\;,
\end{equation}
where all neurons are considered potential source neurons, and hence, $1/N$ is the probability to draw a specific neuron as source. When a neuron emits a spike, it needs to reach on average $K_N$ synapses distributed across $p_\mathrm{target}\,T$ threads. Hence, the fraction of irregular memory access to synapses in the spike-delivery phase is just
\begin{equation}
    f_\mathrm{irr}^\mathrm{conv} = \frac{p_\mathrm{target}\,T}{K_N}
\end{equation}
for the conventional round-robin distribution scheme.

For the structure-aware strategy, we distinguish between the intra-area synapses, which are all hosted by the same MPI process as the source neuron, and the inter-area synapses, which are hosted by the remaining $M-1$ MPI processes. For simplicity, we assume equally sized areas of $N_M$ neurons and equal amounts of $K_N^\mathrm{intra}=K_N^\mathrm{inter}$ intra-area and inter-area targets per neuron. The probability that a neuron has at least one intra-area target on a specific thread on the MPI process hosting the area is given by
\begin{equation}
    p_\mathrm{target}^\mathrm{intra} = 1 - (1 - 1 / N_M)^{N_T K_N^\mathrm{intra}}\;,
\end{equation}
with the probability $1/N_M$ of drawing a specific neuron of the area as source. For inter-area targets, the probability
\begin{equation}
    p_\mathrm{target}^\mathrm{inter} = 1 - (1 - 1 / (N - N_M))^{N_T K_N^\mathrm{inter}}\;,
\end{equation}
considers the remaining $N-N_M$ neurons as potential sources. The fraction of irregular memory access to the intra-area synapses distributed across $p_\mathrm{target}^\mathrm{intra}\,T_M$ threads of the MPI process hosting the area amounts to $p_\mathrm{target}^\mathrm{intra}\,T_M/K_N^\mathrm{intra}$.
For the inter-area synapses distributed across the $p_\mathrm{target}^\mathrm{inter}\,T_M(M-1)$ threads, the fraction of irregular memory access is
$p_\mathrm{target}^\mathrm{inter}T_M(M-1)/K_N^\mathrm{inter}$. Combining the two yields the fraction of irregular memory access to synapses in the spike-delivery phase
\begin{equation}
    f_\mathrm{irr}^\mathrm{struc} = \frac{p_\mathrm{target}^\mathrm{intra}\,T_M+p_\mathrm{target}^\mathrm{inter}\,T_M(M-1)}{K_N}
\end{equation}
for the structure-aware distribution scheme.

A comparison of the fractions of irregular access predicted for the conventional round-robin distribution scheme and the structure-aware scheme in a weak-scaling scenario reveals an advantage of the structure-aware strategy, which grows with the number of MPI processes and threads (\fref{fig:sim-cycle_times_theory_andn_mem_access}b). At $M=16$ MPI processes, the fraction of irregular access to synapses in the spike-delivery phase is still similar for both strategies. However, at $M=32$ MPI processes the structure-aware strategy shows a reduction in irregular access by $12\,\%$ for $T_M=48$ threads and $29\,\%$ for $T_M=128$ threads compared to the conventional strategy. At $M=128$ processes, the theory predicts an even larger reduction by $37\,\%$ for $T_M=48$ and $43\,\%$ for $T_M=128$.

\subsection{Benchmarking results}
\label{sec:Benchmark-results} 
The above theoretical analyses predict a gain in performance for the spike routing in distributed simulations of multi-area models. This applies to both the synchronization required for the regular spike exchange between compute nodes (\sref{sec:synch-time-model}) and the subsequent node-local spike delivery (\sref{sec:memory-access-model}). To evaluate these predictions, we use two network models throughout the benchmarking experiments: the multi-area model of macaque visual cortex (MAM; \citename{Schmidt18_e1006359}, \citeyear*{Schmidt18_e1006359}; \citename{Schmidt18_1409}, \citeyear*{Schmidt18_1409}) and the MAM-benchmark, a simplified variant of the MAM with homogeneous area sizes and constant firing rates designed for controlled scaling experiments (see \sref{sec:benchmark-network-model} for details). To this end, we have incorporated the structure-aware simulation strategy (\sref{sec:structure-aware-simulation-strategy}) in our reference simulation code (\sref{sec:Reference-implementation}), which allows us to compare conventional and structure-aware scheme in a consistent setting.

\subsubsection{Weak scaling}
\label{sec:weak-scaling-results}
\begin{figure}
      \includegraphics[scale=1]{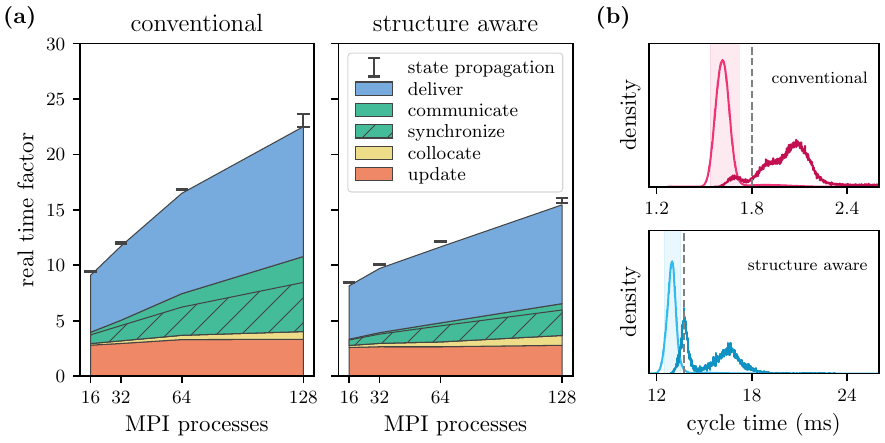}
    \caption{\textbf{Performance comparison of conventional and structure-aware simulation strategy.} Simulations of the MAM-benchmark with overall minimum delay $d_\mathrm{min}=0.1\mathrm{ms}$ and minimum inter-area delay $d^\mathrm{inter}_\mathrm{min}=1\mathrm{ms}$ on SuperMUC-NG using one MPI process and $T_M=48$ threads per compute node. $N_M=130,000$ neurons per process with $K_N=6000$ synapses per neuron are simulated for $T_\mathrm{model}=10\,\mathrm{s}$ biological time. (a) Weak-scaling experiment increasing the number of areas with the number of MPI processes $M$, such that the total number of areas corresponds to $M$. Real time factor as a function of number of MPI processes for the conventional (left) and structure-aware (right) case with contributions of different simulation phases (legend). Black error bars indicate variability across three runs. (b) Distribution of cycle times for the conventional strategy (top, pink) and lumped cycle times for the structure-aware strategy (bottom, cyan) for a simulation with $128$ MPI processes (seed $654$). Arithmetic means are $1.6\,\text{ms}$ (conventional) and $13.0\,\text{ms}$ (structure-aware); shaded bands indicate the standard deviations. Dashed vertical lines mark the upper $3.5\,\%$ quantile each distribution. Lighter curves show the kernel density estimates of all cycle times, darker curves the per-cycle maxima across MPI processes ($D\,=\,10$ times increased horizontal scale for the structure-aware case).}
    \label{fig:weak-scaling}
\end{figure}
First, we analyze the behavior of the two simulation strategies under weak scaling (\fref{fig:weak-scaling}a) using the MAM-benchmark, where the minimum inter-area delay of $d^\mathrm{inter}_\mathrm{min}=1\mathrm{ms}$ is $D=10$ (\eref{eq:d-min-ratio}) times larger than the overall minimum delay of $d_\mathrm{min}=0.1\mathrm{ms}$. The real time factor increases from $9.4$ at $M=16$ MPI processes to $22.7$ at $M=128$ (slope $0.12$) in the conventional case but only from $8.5$ to $15.7$ (slope $0.06$) in the structure-aware case. Hence, the structure-aware approach not only achieves an overall reduction in runtime by up to $30\,\%$ but also shows a more favorable scaling behavior than the conventional strategy. Consistent with the theoretical analysis, the performance improvement is due to speed ups in the spike-delivery and communication phase, where the latter includes both data exchange and synchronization. We observe the largest gain at $M=128$ MPI processes, where spike-delivery time reduces by $25\,\%$ when using the structure-aware instead of the conventional scheme, data-exchange time reduces by $76\,\%$, and synchronization time reduces by $48\,\%$.

The theoretical analysis of spike delivery predicts $37\,\%$ less irregular memory access for a MAM-benchmark simulation with the structure-aware strategy with $M=128$ MPI processes and $T_M=48$ threads per process compared to the conventional approach (\sref{sec:memory-access-model}). The weak-scaling experiments demonstrate that this does not translate one-to-one into a reduction in spike-delivery time. However, as theoretically predicted (\fref{fig:sim-cycle_times_theory_andn_mem_access}b), spike delivery increasingly benefits from the structure-aware approach under weak scaling. In the conventional case, the target synapses of each neuron are ever more distributed across MPI processes and threads, until eventually all targets are fully dispersed. Cache efficiency of spike delivery decreases accordingly. In the structure-aware case, on the other hand, only the inter-area target synapses increasingly spread out under weak scaling, while the intra-area targets remain on the same MPI process. This retains cache efficiency for the intra-area spike delivery, which explains the observed performance advantage over the conventional simulation strategy.

The observed reduction in data-exchange time is not immediately intuitive. Although the structure-aware strategy performs global communication $D=10$ times less often than the conventional approach, each communication operation transmits ten times more data. The MAM-benchmarks are, however, in line with the above \texttt{MPI\_Alltoall()} benchmarks (\fref{fig:mpi_benchmark}) demonstrating that the combined latency and data-exchange time increases sublinearly with the message size. Consequently, sending fewer but larger messages as in the structure-aware strategy is more efficient than sending many smaller ones. The achieved $76\,\%$ reduction in data-exchange time for the MAM benchmark at $M=128$ processes is broadly consistent with the $84\,\%$ reduction predicted from the MPI benchmark.

The theoretical prediction of a $68\,\%$ reduction in synchronization time (\sref{sec:synch-time-model}) also exceeds the measured improvement of $48\,\%$. To explain the deviation from the theory, we inspect the measured cycle times in the simulations of the MAM-benchmark (\fref{fig:weak-scaling}b). The cycle time distributions are shown for a single seed, as \fref{fig:weak-scaling}a exhibits negligible variability across seeds and each run already yields $\sim 10^5$ samples per process, producing stable statistics.
The distribution of measured cycle times in the conventional case is bimodal showing a major-mode peak around $1.62\,\mathrm{ms}$ next to a minor-mode peak around $1.90\,\mathrm{ms}$. The longest recorded cycle time is $18.35\,\mathrm{ms}$. In the structure-aware case, we also observe a bimodal distribution with a major-mode peak at $13.0\,\mathrm{ms}$ and a minor-mode peak at $16.62\,\text{ms}$; the longest cycle time is $29.66\,\mathrm{ms}$. The main body of the structure-aware distribution is shifted by a factor of approximately $8.1$ with respect to the conventional distribution, and the ratio of the respective arithmetic means is $8.0$. This value is smaller than the theoretically predicted shift of $D=10$ (\eref{eq:normal-dist-lumped-cycle-times}), which we attribute to the $24\,\%$ shorter delivery phase in the structure-aware scheme. 
Moreover, as predicted, the relative dispersion from the mean is significantly smaller in the structure-aware case ($CV_\mathrm{struc}=0.040$) than in the conventional case ($CV_\mathrm{conv}=0.056$), which accounts for the observed reduction in synchronization time. However, the ratio of the coefficients of variation of the two distributions ($0.71$) falls between the theoretical bound of $1/\sqrt{D}=0.32$ derived under the assumption of fully independent consecutive cycle times (\eref{eq:rel_dispersion}) and the bound of $1.0$ derived under the assumption of fully correlated consecutive cycle times.
This reflects the presence of partial serial correlations in the cycle times of individual MPI processes. The temporal evolution of per-process cycle times (\fref{fig:cycle_time_heatmap} in appendix) reveals that processes are not only systematically faster or slower on average but also exhibit extended phases of elevated or reduced cycle times that may persist over several thousand cycles.

Based on the measured cycle-time distributions, the theoretical analysis predicts that the upper $99\,\%$ of the per-cycle maxima are located between $1.80\,\mathrm{ms}$ and $18.35\,\mathrm{ms}$ in the conventional case and between $13.70\,\mathrm{ms}$ and $29.66\,\mathrm{ms}$ in the structure-aware case (\eref{eq:prediction-per-cycle-maximum-interval}). In fact, only the upper $91\,\%$ (conventional) and $84\,\%$ (structure aware) fall within the predicted interval. The theory assumes the cycle times of individual processes to be fully uncorrelated. However, this is not the case for the measured cycle times (see \fref{fig:cycle_time_heatmap} in appendix). MPI processes often concertedly take longer or shorter time to go through specific simulation cycles. 

On the other hand, the measured $91\,\%$ (conventional) and $84\,\%$ (structure aware) of the per-cycle maxima do not stray too far from the theoretical $99\,\%$. This means that the corresponding upper $3.5\,\%$ of the measured cycle times do not typically coincide in the same cycles but dominate individual cycles -- otherwise, they would not prevail the per-cycle maxima distributions. The upper $3.5\,\%$ mainly comprise the minor-modes of the recorded cycle times. These large cycle times are unrelated to changes in spiking activity as the MAM-benchmark does not exhibit different activity modes. Unraveling the source of the occasional large cycle times is however not within the scope of this study. Whenever these extremes influence isolated simulation cycles of individual processes, the structure-aware strategy can mitigate their impact on synchronization time.

\subsubsection{Effect of heterogeneity}
\label{sec:benchmarks-heterogeneity}
\begin{figure}
    \centering
    \includegraphics[scale=1]{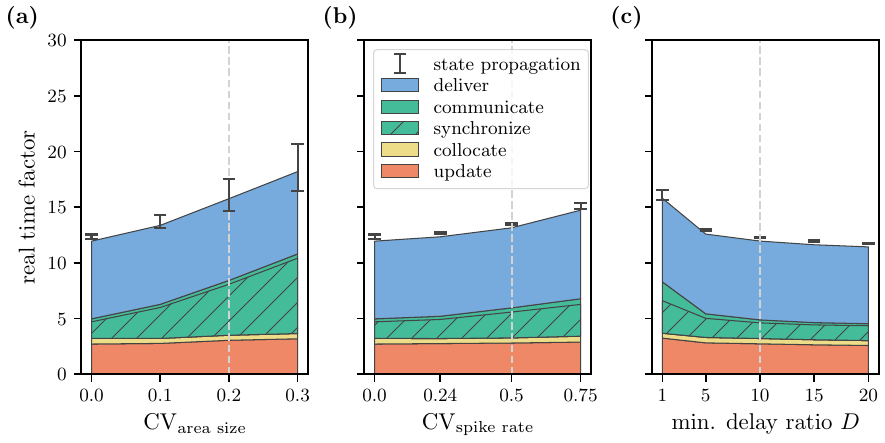}
    \caption{\textbf{Effect of inter-area variability and delay on the performance of the structure-aware simulation strategy}. Simulation of the MAM-benchmark on SuperMUC-NG using $M=64$ MPI processes and compute nodes with one area per process. Color scheme, and benchmarking configurations as in \fref{fig:weak-scaling} unless otherwise stated. Real time factor as a function of (a) variability in area size with a fixed mean area size of $130,000$ neurons, (b) spike-rate variability across areas with a fixed mean rate of $2.5$ spikes per second, and (c) ratio $D$ of minimum inter-area delay to fixed overall minimum delay $d_\mathrm{min}=0.1\,\mathrm{ms}$. Area sizes and spike rates are drawn from normal distributions with coefficients of variation $\mathrm{CV}_\mathrm{area\;size}$ and $\mathrm{CV}_\mathrm{spike\;rate}$, respectively.  Each data point represents the average over three independent sampling seeds. Gray dashed lines indicate the parameter values corresponding to the MAM in \fref{fig:MAM-ori_morph-performance}.}
    \label{fig:params-variation}
\end{figure}
In all of the above simulations, MPI processes handle the same workload in terms of area size and spike rate, not only in the conventional case but also when using the structure-aware strategy. This is due to the homogeneous network configuration. Besides, the minimum delay ratio of $D=10$ imposes a lower bound of $d^\mathrm{inter}_\mathrm{min}=1\mathrm{ms}$ on the delays between areas given the overall minimum delay of $d_\mathrm{min}=0.1\mathrm{ms}$. We now assess the robustness of the structure-aware scheme to heterogeneous conditions. To this end, we analyze the performance of simulations of the MAM-benchmark with a fixed size of $64$ areas using $M=64$ MPI processes and compute nodes for a set of controlled parameter variations (\fref{fig:params-variation}).

In two separate experiments, we investigate the effect of, firstly, different area sizes, and secondly, different per-area spike rates. Area sizes and spike rates are drawn from normal distributions, respectively, where the standard deviations increase but the means are fixed and correspond to the $M=64$ case of the weak-scaling experiments (\fref{fig:weak-scaling}a). To interpret the effect on the different simulation phases, we need to consider that in the reference implementation, synchronization time accounts for the variability across MPI processes (\sref{sec:Reference-implementation}). If one MPI process is systematically slower than the others in the update, collocate, or deliver phase, for example, due to hosting the largest area, it will throughout the simulation make the other processes wait in the synchronization phase.

With increasing variability in area size, the overall runtime also increases (\fref{fig:params-variation}a). The individual update times of the MPI processes correlate with the more and more uneven amounts of local neurons. However, as expected, this imbalance is not reflected by an increase in the averaged update time but in the synchronization time. The measured real time factors also increasingly vary across runs. This is because, for larger standard deviations of the underlying normal distribution, the $64$ sampled area sizes of a specific simulation run are more likely to misrepresent the distribution.

For the investigated regime of low spike rates with less than $10$ spikes per second, a higher spike-rate variability only moderately affects synchronization time and thus overall runtime (\fref{fig:params-variation}b). We infer from this that the MPI processes hosting the more active areas do not spend significantly more time going through a simulation cycle than the processes hosting the less active areas. This means that detecting the spikes of the process-local neurons and buffering them in the spike register are not the most time-consuming steps of the update phase. The collocate phase contributes little to the overall runtime to begin with, such that an elevated local spike rate on some MPI processes can not introduce a major workload imbalance between processes. For the deliver phase, we need to distinguish between the delivery of intra-area spikes and inter-area spikes to their process-local targets. The MPI processes hosting the more active areas have an increased workload compared with the less active ones in the intra-area spike delivery. Yet, the effect on synchronization time is moderate. Intra-area spike delivery is less critical for performance than inter-area spike delivery as the latter is considerably more demanding in terms of memory access (\sref{sec:memory-access-model}). However, as the mean spike rate is fixed in this experiment, the increase in spike-rate variability does not substantially change the total number of spikes that each area receives from the other $63$ areas. This limits the impact on synchronization time.

Finally, in a third experiment, we systematically vary the minimum delay ratio $D$ (\eref{eq:d-min-ratio}) to assess how the performance of the structure-aware scheme depends on the degree of separation between intra- and inter-area delay distributions (\fref{fig:params-variation}c). As the minimum delay is fixed at $d_\mathrm{min}=0.1\,\mathrm{ms}$, increasing $D$ raises the lower cutoff for inter-area delays to $d^\mathrm{inter}_\mathrm{min} = D\cdot d_\mathrm{min}$. This constrains the heterogeneity of inter-area delays, but it allows us to decrease the frequency of global spike communication (\sref{sec:structure-aware-simulation-strategy}). For $D=1$, global spike exchange occurs after every simulation cycle, matching the communication frequency of the conventional scheme. 
Even in the absence of any minimum delay separation, the structure-aware scheme still outperforms the conventional one (see \fref{fig:weak-scaling}a at $64$ MPI processes for reference), as the improved cache efficiency of the structure-aware neuron distribution reduces spike-delivery time regardless of the communication scheme.
Communication time decreases rapidly when increasing the ratio to $D=5$. This is due to a decrease in both synchronization time and data-exchange time. The gain in performance is less pronounced when further increasing the ratio to $D=10$, and it becomes negligible for larger values of $D>10$. The above weak-scaling experiments show that the theoretical analysis (\sref{sec:synch-time-model}) does not allow for exact quantitative predictions of the observed reduction in synchronization time when using the structure-aware simulation strategy instead of the conventional scheme (\sref{sec:weak-scaling-results}). However, the qualitative prediction that the structure-aware scheme is most effective for small minimum delay ratios (\eref{eq:theoretical_ratio_synch_times}) is confirmed by the rapid decrease in synchronization time for small values of $D<10$ observed in this experiment.
Taken together, these results provide insight into the applicability of the structure-aware approach across different network configurations. The communication benefit requires a meaningful separation of delay scales, whereas the delivery benefit of the structure-aware neuron distribution persists even when no such separation exists.

\subsubsection{Real-world network model}
\label{sec:MAM-ori} 
The MAM-benchmark allows us to analyze the performance of multi-area model simulations with the conventional strategy and the structure-aware strategy under controlled conditions. It generates stable workload under weak scaling (\sref{sec:weak-scaling-results}) and enables selective changes to the network configuration to systematically study the effects on different simulation phases (\sref{sec:benchmarks-heterogeneity}).
\begin{figure}
    \centering
    \includegraphics[scale=1]{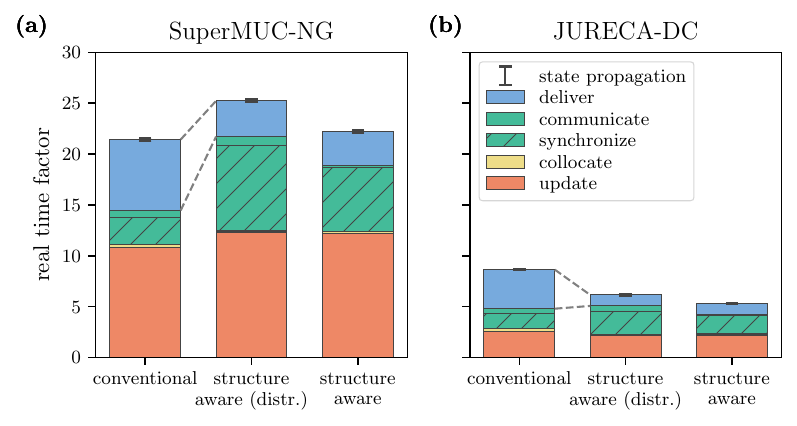}
    \caption{\textbf{Performance comparison of conventional and structure-aware simulation strategy for a real-world network model on two HPC systems.} Simulations of the multi-area model of macaque visual cortex (MAM) in the ground state~ using $M=32$ MPI processes and one process per compute node on (a) SuperMUC-NG with $T_M=48$ threads per process and (b) JURECA-DC with $T_M=128$ per process. Color scheme and black error bars for three runs as in~\fref{fig:weak-scaling}. Left and right bars in each panel show the real time factor for the conventional and the structure-aware strategy, respectively. Middle bars show the real time factors when using a structure-aware distribution scheme for neurons and synapses but conventional global communication every $d_\mathrm{min}=0.1\,\mathrm{ms}$. Dashed lines highlight the change in spike delivery time from the conventional to the structure-aware neuron distribution scheme.}
    \label{fig:MAM-ori_morph-performance}
\end{figure}
To evaluate the potential benefits of the structure-aware approach for day-to-day research, we now compare the two simulation strategies for a real-world network model (\fref{fig:MAM-ori_morph-performance}) -- the multi-area model of macaque visual cortex (MAM). \Sref{sec:benchmark-network-model} summarizes the properties of both MAM and MAM-benchmark.

\begin{figure}
    \centering
    \includegraphics[scale=1]{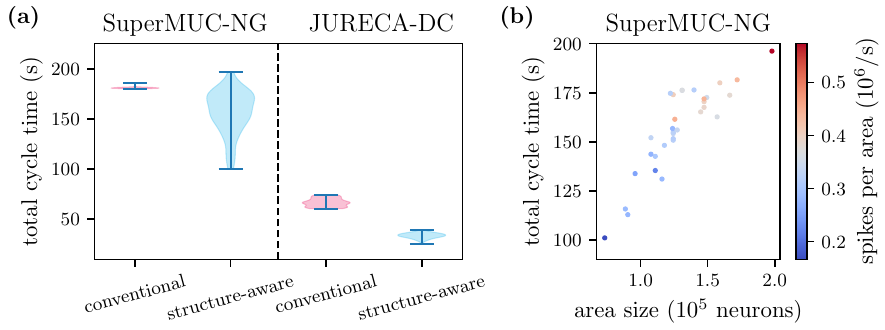}
    \caption{\textbf{Per-process cycle time analysis for simulations of the MAM underlying the results shown in \fref{fig:MAM-ori_morph-performance}.} (a) Violin plots of the distribution of total cycle times across MPI processes for simulations using the conventional (pink) and structure-aware (cyan) strategy on SuperMUC-NG and JURECA-DC. (b) Total cycle time per MPI process as a function of area size for the structure-aware strategy on SuperMUC-NG, with total spike count per area encoded by marker color.}
    \label{fig:per_process_cycle_times}
\end{figure}

We analyze the performance of simulations of the MAM in its ground state on two HPC systems, SuperMUC-NG as well as JURECA-DC (\sref{sec:Benchmark-system}), where besides differences in processor type, available memory, and interconnect, JURECA-DC offers more cores per node than SuperMUC-NG. Next to the conventional and the structure-aware simulation strategy, we test a third strategy, which represents an intermediate step between the two. This intermediate strategy distributes neurons across MPI processes in a structure-aware way (\sref{sec:structure-aware-distribution}) but carries out conventional global communication every $d_\mathrm{min}=0.1\,\mathrm{ms}$. This exposes whether the observed effects of the structure-aware simulation strategy on different simulation phases are due to the neuron-distribution scheme or the reduction in global communication.

On both systems, the structure-aware neuron distribution alone leads to a reduction in spike-delivery time that even exceeds the improvement predicted by the theoretical model for irregular synapse access (\sref{sec:memory-access-model}). This is because the theoretical model assumes a homogeneous connectivity, with equal probabilities for intra- and inter-area connections. In contrast, the MAM is based on experimental connectivity data and exhibits inhomogeneities including a higher proportion of intra-area connections, which amplifies the reduction in irregular synapse access for local connections in structure-aware simulations. The gain in spike-delivery performance, however, is accompanied by an increased synchronization time. As expected from the heterogeneity of the network, there is an unbalanced computational load across the MPI processes. This imbalance is not visible in phases such as delivery, update and collocation, where values are averaged over MPI processes, but manifests directly in the longer synchronization phase. The synchronization phase is reduced again when using the fully structure-aware strategy that separates short and long range communication, yet remains larger than that of the conventional approach.

The high synchronization time is consistent with our observations for the structure-aware scheme on the MAM-benchmark (\fref{fig:params-variation}a), where increasing heterogeneity in area size and spike rate leads to an increase in synchronization. The impact of spike rate heterogeneity is expected to be stronger in the MAM than in the MAM-benchmark (\fref{fig:params-variation}b), as the latter employs a simplified neuron model that produces spikes at a constant rate without propagating membrane potentials (see \sref{sec:benchmark-network-model} for more details). In contrast, the state update of the integrate-and-fire neurons in the MAM is more sensitive to variations in firing activity across areas and over time. 
As a result, both structural heterogeneity (area size) and dynamical heterogeneity (neuronal activity) contribute to the elevated synchronization time observed for the structure-aware scheme in the biologically more realistic model.

On both machines, differences in neuron count and neuronal activity directly translate into varying cycle times of the MPI processes for the structure-aware simulation strategy. 
\Fref{fig:per_process_cycle_times} provides a direct visualisation of this load imbalance at the per-process level. On SuperMUC-NG the distribution of cycle times across MPI processes is substantially broader for the structure-aware scheme than for the conventional scheme, consistent with the larger synchronization times observed on the system (\fref{fig:MAM-ori_morph-performance}a). On JURECA-DC, the distributions of the two schemes are of comparable width, consistent with the similarly small difference in synchronization times observed (\fref{fig:MAM-ori_morph-performance}b). The cycle time of each MPI process on SuperMUC-NG correlates with both the size and the total spike count of the area it hosts (\fref{fig:per_process_cycle_times}b). Since these two quantities are themselves correlated in the MAM --- larger areas naturally produce more spikes in total --- the individual contributions of structural and dynamical heterogeneity to the load imbalance cannot be fully disentangled from this data alone.
The process hosting the largest area with the highest spike count, which is area V1 in the MAM, exhibits the largest average cycle time, whereas the process hosting the smallest area with the fewest spikes shows the shortest. In the MAM, V1 hosts $53\,\%$ more neurons and generates approximately $68\,\%$ more spikes than the network-wide averages. On SuperMUC-NG, this translates into a $24\,\%$ longer average cycle time for the corresponding MPI process compared to the mean across processes, and a $94\,\%$ larger total cycle time for the process hosting the largest area compared to the one hosting the smallest. On JURECA-DC, the same differences in area size and spike count result in only a $7\,\%$ longer cycle time and a $54\,\%$ larger total cycle time difference between the most and least loaded processes. This indicates that JURECA-DC, with its higher core counts and greater per-node computational capacity, is less sensitive to workload imbalance, leading to smaller synchronization penalties.

Overall, the results obtained on JURECA-DC follow the same qualitative trends as on SuperMUC-NG, but simulations are faster due to the larger number of hardware threads per node and the more recent processors. For all simulation strategies, the update, delivery, and communication phases are shorter than on SuperMUC-NG. Spike collocation time remains comparable across both systems, as this phase is executed by the master thread and does not benefit from additional threading. In contrast to SuperMUC-NG, the update phase for the simulations using a structure-aware neuron distribution is faster than for the conventional scheme, suggesting that additional architectural differences, such as cache size or memory bandwidth, influence computation. As predicted by the theoretical analysis in \fref{fig:sim-cycle_times_theory_andn_mem_access}b, the benefit of the structure-aware neuron distribution increases with the number of available hardware threads. Consequently, the reduction in delivery time achieved by the structure-aware strategies is more pronounced on JURECA-DC than on SuperMUC-NG. Together, these effects result in a net speed-up of the fully structure-aware strategy compared to the conventional approach by $42\,\%$.

Overall, the results demonstrate that for a model as unbalanced as the MAM, the fully structure-aware strategy outperforms the conventional approach on high-capacity hardware, while performing comparably when the individual nodes struggle with the computational load of an area.

\FloatBarrier
\section{Discussion}
\label{sec:Discussion}

%
The present work investigates how structural and temporal properties of large-scale network models can be exploited to improve the efficiency and scalability of distributed spiking neuronal network simulations on conventional HPC systems. We show that MPI synchronization dominates spike communication as the number of MPI processes increases. Pure MPI data transfer accounts for only a minor fraction of the overall cost. The work introduces a structure-aware simulation strategy that combines a neuron distribution aligned with the modular organization of multi-area networks and a communication scheme exploiting synaptic transmission delays. Consistent with the proposed theoretical analysis of simulation performance, we find that the new strategy substantially reduces spike delivery and synchronization overhead, with the magnitude of the improvement depending on network heterogeneity and hardware characteristics.

%
Confining brain areas to individual compute nodes is the precondition for exploiting the delays of long-range connections in the communication between nodes. The disadvantage is that this destroys the load balancing of the conventional round-robin distribution over compute nodes. As a consequence, compute nodes hosting brain areas with a larger number of neurons or higher activity have a higher computational load than other compute nodes. In each communication cycle, all nodes have to wait until these heavily loaded nodes have completed their tasks. Thus, when the nodes of an HPC system struggle with the load the neuronal network imposes on cache or memory bandwidth, the additional compute time inflicted by the imbalance may be larger than the gain of the structure-aware communication scheme. A simple scaling argument is sufficient to explain this scenario. If the times for the deliver and update phase are large, also their variability across nodes is large in absolute terms. A small relative increase in imbalance can therefore shadow the gain by fewer synchronizations. The comparison of SuperMUC-NG and JURECA-DC in \fref{fig:MAM-ori_morph-performance} demonstrates the effect. This observation is compatible with the low costs of communication in a prototype multi-GPU code for the same network model \cite{Tiddia22_883333}. Here, bottlenecks are such that the GPUs drain their inputs and finish almost at the same time independent of the respective load.

The MAM is an established model in neuroscience and still a challenging benchmark for most neuromorphic computing systems. Its modular organization, with a separation between local (intra-area) and long-range (inter-area) connectivity, makes it particularly suitable for evaluating structure-aware communication strategies. At the same time, it lies at the current limit of large-scale simulation capabilities, and even modest extensions remain difficult for existing neuromorphic platforms.
Despite its success in simultaneously explaining a range of neuroscientific observations the model exhibits known limitations. In particular,  the meta-stable state most compatible with nature is reached by a certain setting of a coupling parameter that cannot be explained in the scope of the model. While the model nicely reproduces aspects like the distribution of spikes rates and the power spectrum of neuronal activity, the meta-stable state still comes with an excess of spike synchrony on a fine time scale in some cortical areas. The need to occasionally communicate and process a massive amount of synchronous spikes puts an unnatural stress on any computing system. In the structure-aware distribution of neurons this increases the imbalance, while in the round-robin scheme, the synchronous neurons are distributed over all nodes. The present work therefore concentrates on the ground state of the network for the price that it underestimates the natural variability of spiking activity over time. 
While the use of the MAM constrains the range of network architectures considered, it provides a well-established and highly relevant benchmark. Importantly, our results obtained with the more general and abstract version of the MAM --- the MAM-benchmark --- indicate that the proposed approach is not limited to strongly modular systems. Performance gains persist even when the separation between inter- and intra-area delays is reduced, suggesting that improved locality alone can be beneficial. This points toward a broader applicability to networks with weaker structure or more homogeneous delay distributions.
An improved version of the MAM that explains the coupling parameter and removes excess synchrony is presently under investigation \cite{Pronold24_979}. Once established, this model or future variants may replace the original MAM as a benchmark.

%
The MAM-benchmark model used in this study enables systematic evaluation of our algorithm across a wide range of network sizes (\fref{fig:weak-scaling}) while maintaining constant activity levels, as well as controlled parameter variations. Its strong-scaling characteristics are comparable to the MAM under the conventional simulation scheme (see \fref{fig:benchmark-models}). However, the simplified neuron model of the MAM-benchmark differs from standard integrate-and-fire neurons in that its update cost is largely independent of spiking activity. As a consequence, it does not capture workload imbalances arising from heterogeneous firing rates, which significantly influence synchronization and performance in the MAM (\sref{sec:MAM-ori}). In this respect, the MAM-benchmark does not fully expose the performance implications of the structure-aware scheme for more biologically realistic models, although it remains a valuable tool for controlled scalability studies.

%
Previous work has demonstrated that substantial scalability gains can be achieved by tailoring simulation strategies to specific structural properties of neural networks. Tiling-based approaches that enforce strong spatial locality show good weak scaling performance for regular, spatially local networks, but are difficult to generalize to heterogeneous multi-area models with dense long-range connectivity \cite{Kozloski11_5_15,yamaura20_16}. Earlier structure-aware frameworks such as the SPLIT simulator \cite{Hammarlund98_443} showed that exploiting explicit population structure and synaptic delays can also yield efficient parallel execution and was successfully used for neuroscience projects for a number of years. Some of the ideas were later transferred into the MUSIC communication library \cite{Djurfeldt10_43}. More recently, CORTEX \cite{Lyu24} has demonstrated impressive performance at extreme scale by redesigning the simulation framework and close hardware co-design. 
From this embedding of our work into the literature of the last three decades it is apparent that authors constructed algorithms primarily motivated and guided by their own neuroscientific needs and interests. This is good because it ensured that practically relevant solutions emerged. Instruments were not build for their own sake but for achieving concrete scientific insights. These specializations may have hampered the transfer of the technologies into generic simulation code. However, the sustainable development of scientific codes was subject to probably more important sociological and financial problems discussed elsewhere \cite{Aimone23_418,Hocquet24_1,Plesser25_295}. 
The present work takes a middle-ground. Our reference implementation is done in the framework of an established generic simulation code. Therefore researchers can start to experiment with the structure aware assignment of neurons in the environment they are used to and maintaining the full generality of the code. It is our hope that by this we can incrementally make further steps towards a more useful scientific instrument.
Whether partitioning algorithms could automatically discover delay-based modularity in networks where structure is not known a priori is an interesting direction for future work, but lies outside the scope of the present study, which focuses on the biologically relevant case where network structure is defined explicitly during model construction.

%
The impact of our work on neuroscientific investigations will first be limited. Our reference implementation is based on the established simulation code NEST and we do not expect major obstacles by merging the new technology into future releases. Nevertheless, 
NEST places neurons on compute nodes in a round-robin fashion. Therefore, we achieve the mapping of brain areas to compute nodes by respecting the specific numbering of neurons in network construction. This needs to occur in the programmatic model specification on the user level. For a more general application high-level language constructs need to be developed with the concept of an area. In this way, the simulation engine can organize an adequate assignment of neurons in the background.

However, our algorithm already advances CPU reference code for multi-GPU solutions and neuromorphic systems. It separates physical from software limits. The insight that synchronization of compute nodes and not latency is the bottleneck of communication enables the community to have a closer look at synchronization and rethink ideas of asynchronous communication. As recently done for the single compute node \cite{Senk26}, we can begin to explore for multi-node simulations the expected gains of technology scaling versus architectural innovations.

%
Even in the idealized case of equal size of the circuits representing cortical areas and homogeneous and stationary spike rates of the neurons, the time required for the synchronization between compute nodes remains the major contribution to total communication time (\fref{fig:params-variation}). The dominant contribution to state propagation, however, is the phase of spike delivery, and its variability is the origin of what we visualize as synchronization time (\sref{sec:synch-time-model}). Considerable efforts have already been undertaken to streamline the routing of spike to their target neurons \cite{Pronold22_102952,pronold2022}, not all of these ideas have yet entered production code. Previous work has provided a foundation for semi-empirical performance models and scaling analyses of large-scale NEST simulations and storage hierarchies \cite{Schenck14_SC,schenck2017evaluation}. Nevertheless, it seems that it is time for more detailed measurements and more advanced performance modeling to learn about the detailed bottlenecks and create ideas on how to circumvent them. 
The parallel compute power of compute nodes is still increasing. Current high-end compute nodes feature in the range of $256$ computational cores. For a network of $100,000$ neurons distributed across all cores of such a node, each core would only have to take care of around $400$ neurons and consequently oversee only a limited amount of memory. In principle this should remove the von Neumann bottleneck for the intra-node computation. But to what extent can we exploit the microscopic parallelism of spiking neuronal networks with conventional computer architectures? Initial discussions over the past few years have led to the formation of a new consortium to develop respective performance models \cite{Mod4CompWebpage}. 

The present work assigns the network representation of each cortical area to just one compute node. With the help of \texttt{MPI\_Group} future work can relax this constraint to a more flexible scheme where a cortical area extends over multiple compute nodes forming an \texttt{MPI\_Group}. In this way the number of compute nodes per area adapts to the size of the area such that number of neurons per compute node remains approximately constant. Within an \texttt{MPI\_Group}, communication occurs in intervals of the minimal delay of the connections of the area until the minimal delay of the long-range connections is covered and a communication between the areas needs to occur. The scheme regains load balancing with respect to network structure.

A fundamental limitation remains the need to first synchronize threads and then synchronize compute nodes in certain intervals due to the collective communication. Previous studies have shown that the choice of communication mechanism (blocking, non-blocking, or collective) significantly affects scalability in distributed spiking simulations \cite{Thibeault13}. However, even non-blocking approaches do not fully eliminate synchronization constraints at scale. Recent work has improved the interplay between MPI and OpenMP, facilitating asynchronous communication \cite{Protze22}. It remains to be investigated whether spiking network simulation technology can benefit from these improvements, or whether this would require novel hardware support.

%
More powerful compute nodes may even bring real-time performance into sight. As long as computation dominates, mainstream supercomputing infrastructure remains well suited for advancing large-scale brain models. However, if future hardware substantially accelerates per-node computation, communication may again emerge as the limiting factor. In that regime, further progress may require new communication technologies or architectural paradigms beyond today’s standard HPC systems.

\FloatBarrier
\section{Methods}
\label{sec:Methods}

\subsection{Reference implementation}
\label{sec:Reference-implementation}

The neuronal simulation tool NEST version $3.6$~\cite{NEST360} serves as the reference simulation framework for all implementations presented in this study. NEST is an open-source software tool for large-scale networks of spiking neurons, designed to efficiently model neuronal dynamics on computing platforms ranging from desktop systems to high-performance supercomputers. Users control network specification and simulation procedure through a high-level Python module PyNEST \cite{Eppler09_12,Zaytsev14_23_tmp}, while all computationally intensive tasks are executed in a C++ simulation kernel. To exploit multi-core hardware and enable users to run large-scale networks, NEST employs hybrid parallelization, distributing neurons across MPI processes and using multi-threading within each process via OpenMP. The code assigns neurons to threads and MPI processes in a round-robin fashion based on their unique global id (GID) given by order of creation. This distribution scheme ensures that neurons belonging to the same population are evenly distributed across compute nodes. Therefore, the round-robin distribution strategy helps to reduce workload imbalances arising from correlated activity within the populations. 

Simulations advance in discrete cycles that consist of three phases: event delivery to target neurons, neuronal state updates, spike collocation. Following these phases, a global spike communication step exchanges spikes between MPI processes to preserve causality (see \fref{fig:algorithm}). NEST records the wall-clock time of the individual simulation phases using internal high-resolution timers (see NEST documentation\footnote{\url{https://nest-simulator.readthedocs.io/en/stable/index.html}} for further details). For a given simulation cycle $s$ and MPI process $i$, we define cycle time as
\begin{equation}
    T_{s,i} = T_{s,i}^\text{deliver} + T_{s,i}^\text{update} + T_{s,i}^\text{collocate}\, .
\end{equation}
The measure excludes the time required at the end of each cycle for communication between the nodes. As soon as each MPI process completes its local collocate phase, it enters the communication phase. However, spike exchange can only begin after all MPI processes have reached this point. Consequently, faster processes must wait for the slowest process $i_\text{max}$ with the largest cycle time $T_{s,i_\text{max}}$, before the actual spike communication can proceed. We refer to this waiting period as the \textit{synchronization} phase and denote its duration for MPI process $i$ by $T_i^\text{synch}$. Following synchronization, spikes are exchanged via a collective \texttt{MPI\_Alltoall()} operation. To distinguish synchronization overhead from the time spent on the actual data exchange, we insert an explicit \texttt{MPI\_Barrier} immediately in front of the \texttt{MPI\_Alltoall()} call. The synchronization time is then measured as the average time spent by the MPI processes waiting at this barrier. The combined time spent on the \texttt{MPI\_Barrier} and the subsequent \texttt{MPI\_Alltoall()} call is referred to the communication phase with duration $T_i^\text{comm}$. Timers for all phases are advanced in each simulation cycle. For analysis, we average the resulting cumulative phase durations across all MPI processes to obtain the total wall-clock time of the simulation. We quantify simulation performance using the real time factor, defined as wall-clock time normalized by simulated model time.

The communication algorithm sends spikes only to those processes that host at least one of their postsynaptic targets. Target tables on the presynaptic side (see \fref{fig:data-structures}), revealing the MPI processes associated with each neuron's outgoing connections, facilitate this procedure. With the optional \textit{spike compression} feature (NEST $3.0$; \citename{Albers22_837549}, \citeyear*{Albers22_837549}) the algorithm sends spikes to target MPI processes only once as opposed to one per target thread. Spikes carry the information of the source neuron identifier and the time step at which the spike occurred within the cycle. Upon arrival at the target process, the spike uses the identifier to look up the corresponding target synapses and neurons via connection and source tables that are presorted by source neuron. If a process generates more spikes than fit into the current spike buffer, all MPI processes resize their buffers accordingly, followed by a secondary communication round.
Spikes delivered to their target neurons are placed into a neuron-local ring buffer (a circular data structure used to store time-delayed inputs). Each spike remains in the ring buffer for a duration corresponding to its synaptic transmission delay minus the time already elapsed since the source neuron fired. Once this remaining buffering time has expired, the spike is retrieved from the ring buffer and contributes to the state propagation of the target neuron in the update phase of the corresponding simulation cycle.

NEST provides a collection of neuron and synapse models at different levels of biological and computational complexity. Beyond the built-in models, users can extend the simulator by defining custom neuron and synapse models using the domain-specific model description language NESTML \cite{Plotnikov16_93,Linssen25_1544143} or by implementing new models directly in C++.

\subsubsection{Structure-aware distribution of workload}
\label{sec:structure-aware-distribution}
Our reference implementation of the structure-aware distribution of neurons does not modify the round-robin distribution scheme of the reference simulation platform NEST. Instead, we realize all changes required to override the effective neuron placement at the level of the Python user interface. This is possible due to a feature introduced in NEST version $3.0$~\cite{Nest30}, which represents neuron populations as node collection objects with array-like semantics, including indexing, slicing, and element-wise operations.

For multi-area models with homogeneous neuron parameterization, all neurons can be created en bloc in a single \texttt{Create()} call, as opposed to instantiating neurons for each area separately. This yields a single node collection containing all neurons in the model. With detailed knowledge of both the model's area sizes and the way the NEST kernel assigns neurons to MPI processes under the round-robin scheme, areas can then be retrospectively defined such that all neurons belonging to a given area are hosted on one compute node. \Fref{fig:morph} illustrates the effect of this procedure using, as an example, a three-area network instantiated on three MPI processes, each running two threads. A conventional executable model specification expressed in the Python interface of NEST, creates neurons area by area, thereby  distributing areas evenly across the available MPI processes by the round-robin scheme. As a result, each process hosts neurons belonging to multiple areas. The structure-aware distribution scheme creates all neurons at once and distributes them evenly across the MPI processes in exactly the same way as in the conventional scheme, resulting in an identical placement of neuron identifiers. However, in a subsequent step, the model specification assigns the neurons on the same MPI process to the same area by constructing a respective node collection. Thus, the present study does not require changes to the Python interface of NEST, but exploits high-level commands for the manipulation of neuron populations and knowledge of how NEST assigns neurons to compute nodes.

For models with heterogeneous area sizes, this procedure may require creating more neurons than in the standard round-robin scheme. Since the underlying distribution mechanism remains unchanged, NEST still assigns the same number of neurons to each MPI process. To ensure that the largest area can be placed entirely on a single process, the total number of neurons must therefore be determined by the maximum area size. As a consequence, MPI processes hosting smaller areas contain additional "ghost neurons", which are explicitly tagged as "frozen" and do not take part in the simulation. The number of synapses is not affected by this procedure.

The structure-aware distribution scheme can be applied using standard NEST without modifications to internal data structures and communication mechanisms. Using NEST version $3.6$, we evaluated the performance impact of the resulting load imbalance for simulations of heterogeneous multi-area models on simulation phases other than communication, such as spike delivery. As expected, the tests revealed degradation in the communication phase, but also showed promising results for the other simulation phases. Based on these findings, we concluded that changes to the kernel code targeting spike communication (described in the following section) could further improve overall performance.

\subsubsection{Two separate communication pathways}
\label{sec:two-separate-communication-pathways}
$\quad$\\
\begin{figure}
    \centering
    \includegraphics[width=\textwidth]{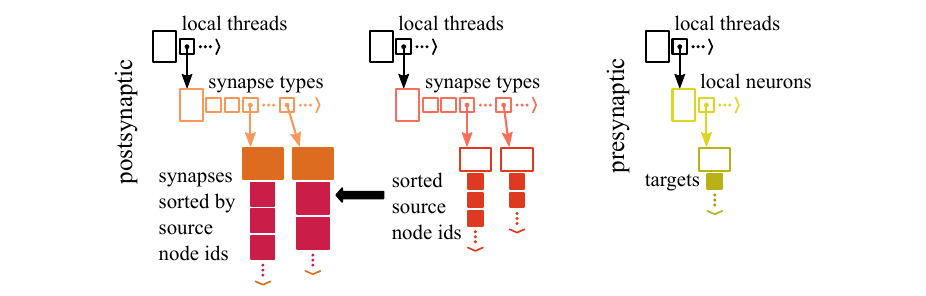}
    \caption{\textbf{Fundamental data structures of the reference code NEST.} Data structures required per MPI process to store the local connections and enable delivery of spikes to local target connections: connection table (left), source table (middle), and target table (right). The outermost dimension of each table is a resizable array of pointers (black) to thread-specific inner resizable arrays. On the receiver side (postsynaptic), the inner resizable arrays of the connection table and the source table store the thread-local connections (pink filled squares) and the corresponding node ids of the presynaptic neurons (red filled squares) sorted by synapse type. On the sender side (presynaptic), the target table stores the locations of the target connections (green filled squares) for every thread-local neuron. Locations are defined by the MPI ranks and the positions in the connection table. Adapted from \protect\citeasnoun{Jordan18_2}, Figure 4A.}
    \label{fig:data-structures}
\end{figure}

To exploit the structure-aware neuron distribution described in the previous section, short- and long-range connections must be explicitly distinguished during the first two main phases of the simulation, network construction and simulation preparation (described below), as they are handled differently in the third main phase: state propagation.

Network construction primarily involves the creation of neurons followed by the creation of synaptic connections between them. Since structure-aware neuron creation itself does not require infrastructural changes, the first kernel-level modifications introduced by the structure-aware NEST implementation arise during synapse creation. At this stage, synaptic data structures such as connection and source tables (see \fref{fig:data-structures}) are duplicated so that connections and their source neuron identifiers are stored separately for short- and long-range connections.
The \texttt{Connect()} call at the Python user interface level accepts a Boolean parameter "long\_range" which is used to assign the  newly created connection to the respective data structure.

While duplicating data structures may raise concerns about increased memory consumption, the additional overhead is small. The outer dimensions of the connection and source tables are defined by thread identifiers and synapse-type identifiers, and the actual connection objects and source-neuron identifiers are not duplicated. Instead, they are distributed across two separate data structures. This results in two shorter data lists as opposed to a single longer one, which may additionally improve cache efficiency during spike delivery.

Following network construction, the simulation enters a preparation phase. This phase first jointly sorts the connection and source tables by source neuron identifiers. Subsequently, it creates a target table that defines the MPI processes hosting the postsynaptic targets of each local neuron and thread (see \fref{fig:data-structures}). The structure-aware strategy again duplicates this target table to mirror the separation of postsynaptic connection and source tables into short- and long-range components. Target table construction thus requires two construction steps, each involving an \texttt{MPI\_Alltoall()} communication of the relevant connectivity data, potentially performed in multiple rounds. 
As with the duplicated connection and source tables, the additional memory overhead due to duplicating the target tables is small, as the algorithm does not replicate the target locations but rather distributes them across the two target tables. 

Finally, the third main phase begins: the state propagation phase advances the dynamical state of the neuronal network. This main phase of the simulation subdivides into the minor phases update, collocate, communicate, and deliver. The structure-aware scheme, excludes neurons tagged as "frozen" on a given MPI process from the update phase (see \fref{fig:algorithm}). 
When a neuron crosses the firing threshold, the algorithm temporarily stores outgoing spikes in a spike register, separately for short- and long-range connections. The structure-aware implementation thus duplicates the data structure. The scheme distinguishes between a local exchange of spikes within the same MPI process and a global exchange of spikes across all MPI processes. Spikes destined for short-range targets are communicated locally and cleared from the corresponding spike register after each exchange. Spikes going to targets of long-range connection are accumulated on the presynaptic side over multiple simulation cycles before being exchanged in the global communication step. The number of cycles over which these spikes are buffered is determined by the minimum delay of long-range connections in the network. Both communication pathways have separate MPI buffers. Since neurons typically exhibit both, short- and long-range outgoing connections, the algorithm inserts their spikes with high probability into both spike registers and MPI buffers. This leads to an increase in memory usage in the structure-aware scheme. Further, buffering long-range spikes over several simulation cycles temporarily raises memory demand on the presynaptic side and results in larger message sizes in the global MPI communication. Once a spike arrives at the target compute node, the separate connection and source tables for short- and long-range connections ensure correct delivery to the postsynaptic neurons, without the need for further duplication of data structures on the postsynaptic side.

\FloatBarrier
\subsection{Benchmark model}
\label{sec:benchmark-network-model}

\begin{figure}
        \centering
        \includegraphics[scale=1.0]{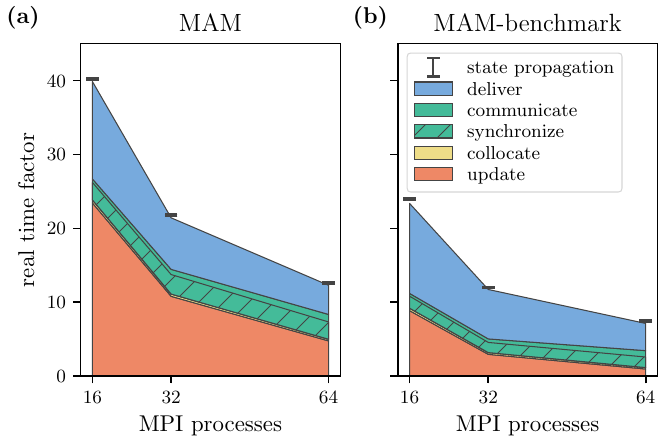}
    \caption{Comparison of strong scaling performance of MAM (a) and MAM-benchmark (b) on SuperMUC-NG, modeling $32$ areas. Color scheme, black error bars and benchmarking configurations as in \fref{fig:weak-scaling}.}
    \label{fig:benchmark-models}
\end{figure}

In this study, we evaluate performance using two spiking neuronal network models. The first is the multi-area model of macaque visual cortex (MAM; \citename{Schmidt18_e1006359}, \citeyear*{Schmidt18_e1006359}; \citename{Schmidt18_1409}, \citeyear*{Schmidt18_1409}). The second is a simplified variant of this model, referred to as the MAM-benchmark, which is designed for scaling experiments and controlled parameter variations.

The MAM comprises one square millimeter of each of the $32$ visual cortical areas of macaque monkey. A layered cortical microcircuit \cite{Potjans14_785} represents the individual areas. Each layer consists of an excitatory and an inhibitory population of integrate-and-fire model neurons all with identical intrinsic parameters. 
Neuron densities and connection probabilities vary across areas and are derived from experimental data, resulting in across-area heterogeneity in both size and spiking activity. The coefficient of variation of area size is $\sim{}0.2$ with an average area size of $130,000$.
The areas are highly interconnected, with approximately a third of all synaptic connections targeting neurons in other areas ($\sim{}1800$ long-range connections per neuron). Synaptic transmission delays are drawn from Gaussian distributions. 
In contrast to the original model we impose a lower cutoff $d_\mathrm{min}^{inter}$ on the delay distribution of the inter-area connections. For the range of $d_\mathrm{min}^{inter}$ values considered in this study, we observe no noticeable effect on network dynamics \cite{Grundler25} although approximately $16\,\%$ of all inter-area connections have a delay below $2\,\mathrm{ms}$.
There exist two dynamical regimes in the MAM depending on its parameterization: a ground state characterized by low and stable firing rates; and a metastable state with episodes of high synchrony and elevated firing rates. The present work considers the ground state with a mean firing rate of $2.5$ spikes per second and a coefficient of variation per time step of $\sim{}0.5$ spikes per second. 

The MAM-benchmark is inspired by the MAM but deliberately more homogeneous. All areas contain the same number of neurons, and each neuron has the same number of intra- and inter-area connections. These numbers are chosen to reflect the average neuron count and connectivity of the MAM across areas, resulting in approximately $130,000$ neurons per area and $6,000$ outgoing connections per neuron, with half of the connections targeting neurons within the same area and half targeting neurons in other areas. Intra-area and inter-area delays are drawn from Gaussian distributions with a mean of $1.25\,\text{ms}$ and $5\,\text{ms}$, and a standard deviation of $0.625\,\text{ms}$ and $2.5\,\text{ms}$, respectively. As in the MAM, we impose a lower cutoff $d_\mathrm{min}^{inter}$ on inter-area delays.
A key difference between the two models lies in the neuron dynamics. The MAM-benchmark employs a single neuron type, referred to as the \textit{ignore-and-fire} neuron. While this neuron receives and emits spikes in the same way as a standard integrate-and-fire neuron, it does not propagate a membrane potential. Instead, it generates spikes at a predefined interval and phase, independent of synaptic input. This simplified neuron model enables us to scale the benchmark-MAM to smaller or larger numbers of areas while maintaining constant activity levels, without the need for retuning network parameters. 

\Fref{fig:benchmark-models} compares the strong-scaling performance of the MAM and the MAM-benchmark. The average delivery, communication and collocation times are very similar for both models, indicating that they exhibit comparable performance-relevant characteristics. As expected, the update phase is noticeably faster for the MAM-benchmark due to simpler neuron dynamics. 

We perform each simulation throughout the study with three different random seeds $\{12,654,91856\}$ for the NEST random number generator. Because connectivity generation in NEST is stochastic and relies on this generator, each seed produces a distinct network instantiation with different connectivity realizations.

\FloatBarrier
\subsection{Benchmarking systems}
\label{sec:Benchmark-system}

We perform benchmarks on two HPC systems: JURECA-DC at the J\"ulich Research Centre, Germany, and SuperMUC-NG at the Leibniz Supercomputing Centre (LRZ) in Garching, Germany. Automatic NUMA balancing by the Linux kernel is set to off in both cases.

The JURECA system \cite{Thoernig21_182} comprises a CPU-only partition (JURECA-DC), as well as a CPU-GPU partition (JURECA-DC Booster). In this study, we use the standard compute nodes of JURECA-DC Phase $2$ ($480$ standard nodes in total). Each node is equipped with two AMD EPYC $7742$ processors, with $64$ cores each, clocked at $2.25\,\mathrm{GHz}$. In total the nodes have $128$ cores and a main memory of $512\,\mathrm{GiB}$. The CPU partition has a peak performance of $3.54$ Petaflops per second. Nodes are interconnected via Mellanox HDR100 InfiniBand. Applications are compiled using the GNU Compiler Collection (GCC), and we link against the OpenMPI library for distributed-memory communication and OpenMP for shared-memory parallelism within a node. We use the general-purpose memory allocator jemalloc version 5.3.0.

SuperMUC-NG Phase 1 \cite{Bastian2024} consists of $6,480$ compute nodes, each equipped with two Intel Skylake Xeon Platinum $8174$ processors running at $3.1\,\mathrm{GHz}$, providing $48$ cores per node. The majority of compute nodes have $96\,\mathrm{GiB}$ of main memory. The peak performance is $26.3$ Petaflops per second. Compute nodes are connected via an Intel OmniPath interconnect and are bundled into eight domains, each employing a fat-tree topology. As on JURECA-DC, we compile applications via GCC and use hybrid parallelism via OpenMP and OpenMPI.

On both systems, we employ one MPI process per node using all available physical cores per node. Hyperthreading is turned off.  

Benchmarks on JURECA-DC use the automated benchmarking pipeline \textit{CI-beNNch} \cite{Vogelsang26_arxiv}, which builds upon the benchmarking framework \textit{beNNch} discussed in \cite{Albers22_837549}. In contrast to \textit{beNNch}, \textit{CI-beNNch} integrates principles from continuous integration, enabling fully automated and user-independent benchmarking workflows. The framework simplifies collaboration among researchers, and promotes reuse and comparison of benchmarking results.

\section*{Acknowledgments}
This project received funding from NeuroSys as part of the initiative “Clusters4Future” funded by the Federal Ministry of Education and Research BMBF (03ZU1106CB,03ZU2106CB); the German Research Foundation (DFG) - 368482240/GRK2416 (MultiSenses-MultiScales) and 545776403/FOR5880 (Mod4Comp); Joint Lab HiRSE, the Helmholtz Platform for Research Software Engineering - an innovation pool project of the Helmholtz Association; the European Union’s Horizon Europe Programme under the Specific Grant Agreement No. 101147319 (EBRAINS 2.0 Project); the Joint Lab ‘Supercomputing and Modeling for the Human Brain’ (SMHB) of the Helmholtz Association; the Volkswagen Foundation; the European Union’s Horizon 2020 Framework Programme for Research and Innovation under Specific Grant Agreement No. 945539 (Human Brain Project SGA3). Melissa Lober received funding from the HIDA-NORA Mobility Program for a research stay at NMBU, Norway.

We gratefully acknowledge the Gauss Centre for Supercomputing e.V. (\url{www.gauss-centre.eu}) for funding this project by providing computing time on the GCS Supercomputer SuperMUC-NG at Leibniz Supercomputing Centre (\url{www.lrz.de}), and computing time granted by the JARA Vergabegremium and provided on the JARA Partition part of the supercomputer JURECA at Forschungszentrum Jülich (computation grant JINB33).

We thank members of the Institute of Advanced Simulation (IAS-6) and the Juelich Research Centre (JSC) at Juelich for their time and immense help in tracking down the unexpected NUMA behavior on JURECA-DC (ticket \#10100311), in particular: Sebastian Gillessen, Alp Inangu, Gorka Peraza Coppola, Hans Ekkehard Plesser, Benedikt Steinbusch, Dennis Terhorst, Guido Trensch, Jan Vogelsang, and Brian Wylie. What a joy to work with such a great team.

\section*{Conflict of Interest Statement} 
The authors declare that the research was conducted in the absence of any commercial or financial relationships that could be construed as a potential conflict of interest.

\section*{Author Contributions} 
All authors listed, have made substantial, direct and intellectual contribution to the work, and approved it for publication.

\section*{Data availability statement}
Data is openly available \cite{Lober26_zenodo}.

\section*{References}
\bibliographystyle{dcu_harvardetal}
\bibliography{bib/add_references, bib/brain}

\clearpage

\section*{Appendix}
\begin{figure}[H]
    \centering
    \includegraphics[scale=0.9]{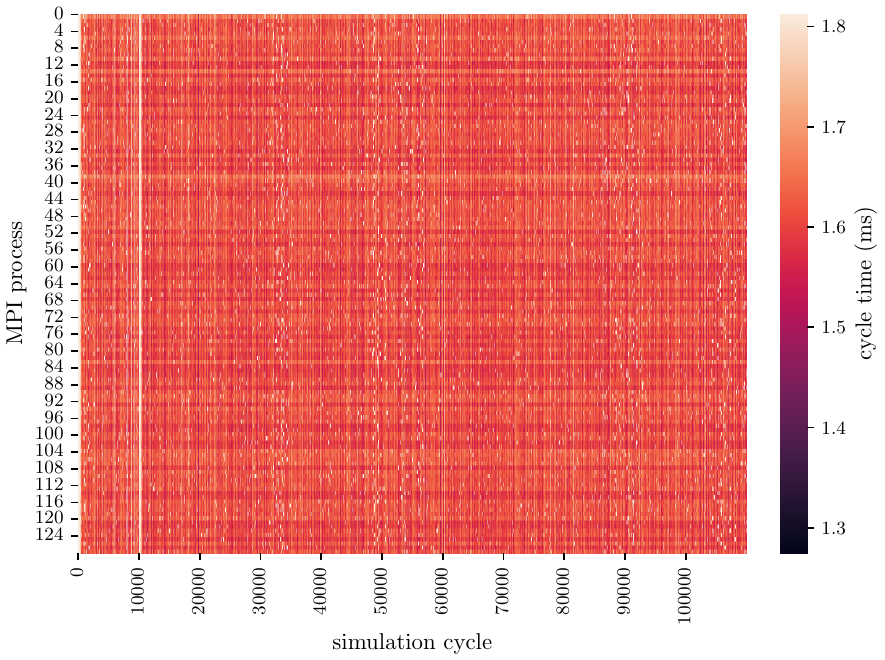}
    \includegraphics[scale=0.9]{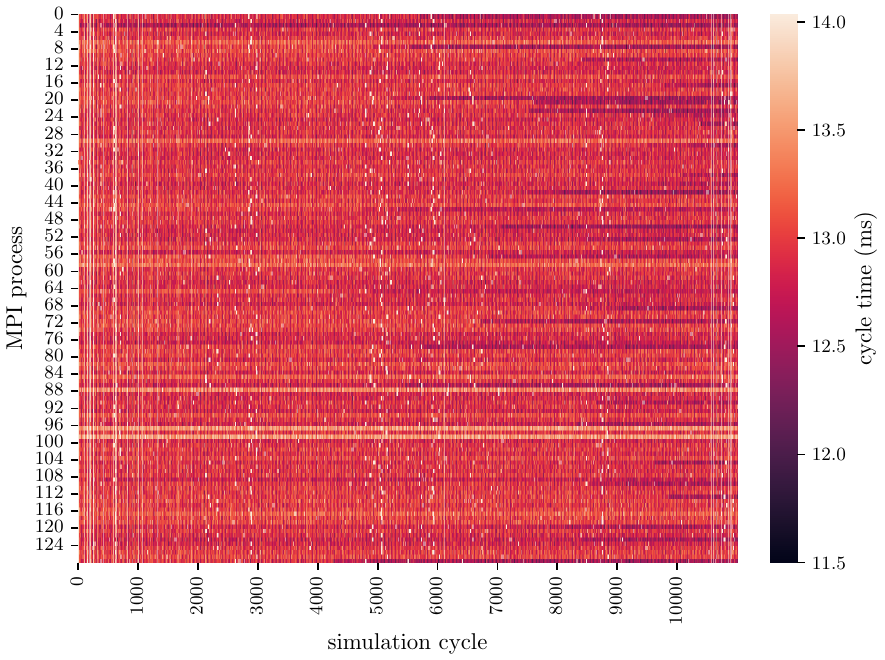}
    \caption{Temporal evolution of the cycle time during a MAM-benchmark simulation with $128$ MPI processes (seed $654$) on SuperMUC-NG, for the conventional (top) and the structure-aware strategy (bottom).}  \label{fig:cycle_time_heatmap}
\end{figure}

\end{document}